\documentclass[twocolumn,astrosymb,twocolappendix]{aastex631}

\usepackage{tablefootnote}

\usepackage[T1]{fontenc}
\usepackage{amsmath}
\usepackage{epigraph}
\usepackage{threeparttable}
\usepackage{soul}
\usepackage{hyperref}

\newcommand{\hide}[1]{}
\newcommand{\angstrom}{\textup{\AA}}

\def\f17{f_{17}}

\def\arcsec{\hbox{arcsec}}

\def\ergcm2s{\ifmmode {\rm\,erg\,cm^{-2}\,s^{-1}}\else
                ${\rm\,ergs\,cm^{-2}\,s^{-1}}$\fi}
\def\ergsec{\ifmmode {\rm\,erg\,s^{-1}}\else
                ${\rm\,ergs\,s^{-1}}$\fi}
\def\mstar{\ifmmode {M^*_{UV}}\else
                ${M^*_{UV}}$\fi}
\def\phistar{\ifmmode {\phi^*}\else
                ${\phi^*}$\fi}
\def\zo{\ifmmode {12 + \log[{\rm O}/{\rm H}]}\else  
                ${12 + \log[{\rm O}/{\rm H}]}$\fi}
                
\def\lya{\ifmmode {\hbox{Ly}\alpha}\else  
                Ly$\alpha$\fi}

\def\OIII{[\mbox{O\,{\sc iii}}]}

\def\OII{[\mbox{O\,{\sc ii}}]}

\def\Lyaesc{$f_{esc}$(Ly$\alpha$)}
\def\Lya{Ly$\alpha$}
\def\Hb{H$\beta$}
\def\HII{\mbox{H\,{\sc ii}}}
\def\HI{\mbox{H\,{\sc i}}}

\shorttitle{Very Blue UV Slope in LyC Leakage}
\shortauthors{Kim et al.}

\begin{document}

\title{Small Region, Big Impact: Highly Anisotropic Lyman-continuum Escape from a Compact Starburst Region with Extreme Physical Properties}

\correspondingauthor{Keunho J. Kim}
\author[0000-0001-6505-0293]{Keunho J. Kim}
\affil{Department of Physics, University of Cincinnati, Cincinnati, OH
45221, USA}
\email{kim2k8@ucmail.uc.edu}

\author[0000-0003-1074-4807]{Matthew B. Bayliss}
\affiliation{Department of Physics, University of Cincinnati, Cincinnati, OH 45221, USA}

\author[0000-0002-7627-6551]{Jane R. Rigby}
\affiliation{Observational Cosmology Lab, Code 665, NASA Goddard Space Flight Center, 8800 Greenbelt Rd., Greenbelt, MD 20771, USA}

\author[0000-0003-1370-5010]{Michael D. Gladders}
\affiliation{Department of Astronomy and Astrophysics, University of Chicago, 5640 South Ellis Avenue, Chicago, IL 60637, USA}
\affiliation{Kavli Institute for Cosmological Physics, University of Chicago, 5640 South Ellis Avenue, Chicago, IL 60637, USA}

\author[0000-0002-0302-2577]{John Chisholm}
\affiliation{Department of Astronomy, The University of Texas at Austin, 2515 Speedway, Stop C1400, Austin, TX 78712, USA}

\author[0000-0002-7559-0864]{Keren Sharon}
\affiliation{Department of Astronomy, University of Michigan, 1085 S. University Ave, Ann Arbor, MI 48109, USA}

\author[0000-0003-2200-5606]{H{\aa}kon Dahle}
\affiliation{Institute of Theoretical Astrophysics, University of  Oslo,  P. O. Box 1029, Blindern, N-0315 Oslo, Norway }

\author[0000-0002-9204-3256]{T. Emil Rivera-Thorsen}
\affiliation{The Oskar Klein Centre, Department of Astronomy, Stockholm University, AlbaNova, SE-10691 Stockholm, Sweden}

\author[0000-0001-5097-6755]{Michael K. Florian}
\affiliation{Steward Observatory, University of Arizona, 933 North Cherry Ave., Tucson, AZ 85721, USA}

\author[0000-0002-3475-7648]{Gourav Khullar}
\affiliation{Department of Physics and Astronomy and PITT PACC, University of Pittsburgh, Pittsburgh, PA 15260, USA}

\author[0000-0003-3266-2001]{Guillaume Mahler}
\affiliation{Institute for Computational Cosmology, Durham University, South Road, Durham DH1 3LE, UK}
\affiliation{Centre for Extragalactic Astronomy, Durham University, South Road, Durham DH1 3LE, UK}


\author[0000-0003-0094-6827]{Ramesh Mainali}
\affiliation{Observational Cosmology Lab, Code 665, NASA Goddard Space Flight Center, 8800 Greenbelt Rd., Greenbelt, MD 20771, USA}

\author[0000-0003-4470-1696]{Kate A. Napier}
\affiliation{Department of Astronomy, University of Michigan, 1085 S. University Ave, Ann Arbor, MI 48109, USA}

\author[0000-0001-7548-0473]{Alexander Navarre}
\affiliation{Department of Physics, University of Cincinnati, Cincinnati, OH 45221, USA}

\author[0000-0002-2862-307X]{M. Riley Owens}
\affiliation{Department of Physics, University of Cincinnati, Cincinnati, OH 45221, USA}

\author[0000-0002-0975-623X]{Joshua Roberson}
\affiliation{Department of Physics, University of Cincinnati, Cincinnati, OH 45221, USA}



\begin{abstract}
\noindent
Extreme, young stellar populations are considered the primary contributor to cosmic re-ionization. How Lyman-continuum (LyC) escapes these galaxies remains highly elusive, and it is challenging to observe this process in actual LyC emitters without resolving the relevant physical scales. We investigate the Sunburst Arc: a strongly lensed, LyC emitter at $z =2.37$ that reveals an exceptionally small scale (tens of parsecs) region of high LyC escape.
The small ($<$ 100 pc) LyC leaking region has extreme properties: a very blue UV slope ($\beta = -2.9 \pm 0.1$), high ionization state (\OIII $\lambda 5007$/\OII $\lambda 3727$ \ $= 11 \pm 3$ and \OIII $\lambda 5007$/\Hb\ $=6.8 \pm 0.4$), strong oxygen emission (EW(\OIII ) $= 1095 \pm 40 \ \angstrom$), and high Lyman-$\alpha$ escape fraction ($0.3 \pm 0.03$), none of which are found in non-leaking regions of the galaxy.
The leaking region's UV slope is consistent with approximately ``pure'' stellar light that is minimally contaminated by surrounding nebular continuum emission or extinguished by dust.
These results suggest a highly anisotropic LyC escape process such that LyC is produced and escapes from a small, extreme starburst region where the stellar feedback from an ionizing star cluster creates one or more ``pencil beam'' channels in the surrounding gas through which LyC can directly escape. Such anisotropic escape processes imply that random sightline effects drive the significant scatters between measurements of galaxy properties and LyC escape fraction, and that strong lensing is a critical tool for resolving the processes that regulate the ionizing budget of galaxies for re-ionization.
\end{abstract}



\section{Introduction}
\label{sec:Sec_1_introduction}
Cosmic reionization is the last major phase transition of the Universe; when most of the neutral hydrogen (\HI ) in the Intergalactic Medium (IGM) became ionized.
Our current understanding of the luminosity functions and Lyman continuum (LyC) escape fractions of AGN and star-forming galaxies suggests that low-metallicity star-forming galaxies are likely the dominant contributors to the reionization process \citep{fan06,robe15,fink19,naid20,yung20}.

However, the escape process of LyC radiation is complex and only a small fraction of star-forming galaxies are confirmed LyC leakers.
This strongly suggests that the escape process crucially depends on the geometry of interstellar medium (ISM), dust screening effects, and the properties of the ionizing stars \citep{zack13,verh15,chis19}. 

What makes a galaxy a LyC emitter?
Over a wide range of redshift ($0.02 < z < 4$), LyC emitters typically have young ($< 10$ Myr) stellar populations, low metallicity (12+$\rm{log(O/H)} < 8.5$), extreme nebular emission line ratios (notably optical \OIII $\lambda 5007$/\OII $\lambda 3727 > 5$), and little dust ($E(B-V) < 0.2$) \citep{berg06,leit13,bort14,most15,leit16,izot16a,izot16,izot18a,izot18,shap16,rutk16,rutk17,vanz18,wang19,malk21,flur22,chis22,marq22}.
They also show a strong \lya\ emission line in their UV spectrum, which suggests a low column density of \HI\ in the ISM and a favorable geometry for the escape of both \Lya\ and LyC photons \citep[e.g.,][]{verh15,rive17}. In particular, there is recent evidence that LyC escape is highly anisotropic, with complex ISM geometries resulting in high LyC escape along a few narrow lines of sight with small solid angles \citep{rive19,ramam20,gaza20}. Directly observing such ``pencil beam'' channels of LyC requires extremely high angular resolution observations of known LyC leaking galaxies.

Notably, LyC emitters show compact morphology with concentrated star formation, as indicated by high star formation rate surface density ($\Sigma$SFR $> 1 \ M_{\odot}\rm{yr^{-1}kpc^{-2}}$) \citep{berg06,bort14,izot16a,izot16,izot18a,izot18,wang19,ji20,kim20,kim21,flur22}.  
Such compact morphology of LyC emitters is closely related to the unresolved star cluster-like compact star-forming regions shown in their UV-continuum images.

Although these properties of LyC emitters show the overall \textit{galaxy} properties, understanding the detailed LyC escape mechanisms---that is, where in a galaxy LyC radiation originates and how it escapes---crucially requires clear spatial information about the leaking galaxy to nail down the distribution of ionizing stars and the geometry of surrounding nebular gas.

To date, such detailed morphologies ($< 100$ pc scale) of LyC emitters have only been obtained for one LyC emitter (aka, Sunburst Arc) at $z=2.37$, that is strongly-lensed by a foreground galaxy cluster at $z=0.44$ \citep{dahl16}. This is because the sub-galactic scale analysis is only possible in strongly lensed LyC leaking galaxies like the Sunburst Arc due to lensing magnification.
We cannot even spatially-resolve the LyC leaking regions of local ($z \sim 0$) LyC emitters because the only instrument with the ability to measure their ionizing photons is the Cosmic Origins Spectrograph on HST, which has a spectroscopic aperture diameter of 2\farcs5.
This means that LyC photons measured from COS cannot be localized on scales smaller than the COS aperture. As an example, for a local LyC leaking galaxy Haro 11 \citep{berg06,leit11,ostl21} at a redshift of $z = 0.02$ \citep{berg86,berg06}, the COS aperture corresponds to a region $\sim$1 kpc in diameter (see Figure 1 of \cite{ostl21} for details).

Unlensed distant (2.5 $\lesssim z \lesssim 3.5$) LyC leakers are observable with HST WFC3/UVIS, but at these cosmological distances even HST is limited to spatial scales of $\gtrsim$ 0.4 kpc. Strongly lensed LyC leaking galaxies like the Sunburst Arc uniquely enable spatially resolved direct studies of LyC escape on the physical scales of individual star clusters. 

Thus, the most detailed spatial information about LyC emitters may be obtained from the Sunburst Arc. Due to lensing magnification, the galaxy's stretched rest-frame LyC image reveals that only one particular star-forming region shows escaping ionizing radiation while other regions within the galaxy do not \citep{rive19}.
Indeed, due to its uniqueness as a bright lensing-magnified LyC emitter, the Sunburst Arc has been of great interest in numerous studies concerning the physics of ionizing radiation production and escape \citep{rive17,rive19,chis19,vanz22,shar22,main22,pasc23,mest23} since its discovery \citep{dahl16}.

In this Letter, we aim to provide the \textit{clearest} view thus far of the physical conditions of a LyC leaking region, by measuring the key physical properties of the Sunburst Arc on exceptionally small scales $<$ 100 pc.
By systematically comparing the physical properties of the leaking region with the non-leaking regions, we investigate whether the leaking region shows any distinct properties that might locally facilitate the escape of LyC photons.
This detailed analysis is made possible through a unique combination of HST's sharp imaging and strong gravitational lensing of this exceptionally bright (integrated $m_{\rm AB} \simeq 17.5$) LyC leaking galaxy.

Section \ref{sec:Sec_2_data_sets} describes the observational data sets and the measurements of the physical properties (UV-continuum slope, ionization parameter, \lya\ escape fraction, and equivalent width of emission lines). We present our results in Section \ref{sec:Sec_3_results} and discuss them in Section \ref{sec:Sec4_discussion}. We summarize our conclusions in Section \ref{sec:Sec_5_Summary_conclusions}. We adopt the $\Lambda$CDM cosmology of ($H_{0}$, $\Omega_{m}$, $\Omega_{\Lambda}$) = (70 $\rm{km\ s^{-1}}$ $\rm{Mpc^{-1}}$, 0.3, 0.7) throughout the paper.

\section{Observations and Data Analysis}
\label{sec:Sec_2_data_sets}

\begin{figure*}
\includegraphics[width=\linewidth]{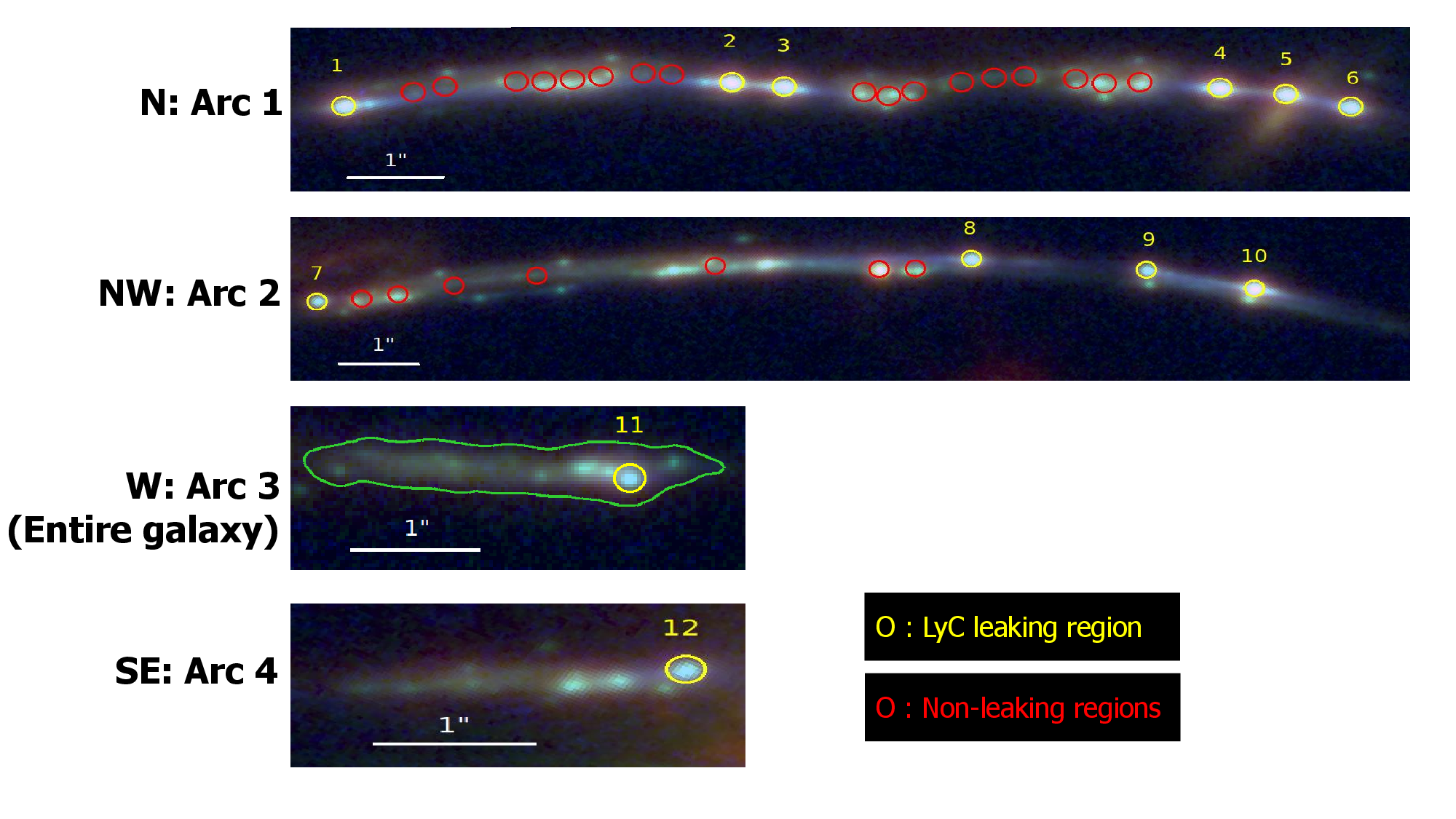}
\caption{HST color-composite images of the strongly lensed LyC galaxy (Sunburst Arc at $z=2.37$), showing its spatially-resolved morphology. 
The image filters (Red: F140W; Green: F606W; Blue: F410M) trace the rest-frame optical, UV-continuum, and \lya\ emission, respectively.
The galaxy appears as four Arcs due to the strong lensing effect by the foreground galaxy cluster PSZ1 G311.65--18.48 at $z=0.44$ \citep{dahl16,shar22}.
Across the arcs, a LyC leaking starburst region appears multiple times, as marked with yellow circles (i.e., 1 to 12).
The lensed images of the leaking region have different lensing magnifications as computed from the lens model \citep{shar22}. Each image also has a slightly different viewing angle from us.
The non-leaking regions used in our analysis are marked as red circles in the Arc 1 and 2 images.
We also measure the galaxy-integrated physical properties from the entire galaxy region outlined in green in the Arc 3 image.
\label{fig0}}
\end{figure*}

\subsection{HST Broadband and Narrowband Imaging}
\label{sec:Section2_1_HST_broad_narrow_imaging}

We analyze HST imaging of the Sunburst Arc taken as a part of several programs: GO-15101 (PI: Dahle), GO-15418 (PI: Dahle), GO-15377 (PI: Bayliss), and GO-15949 (PI: Gladders). Our analysis uses standard \texttt{Astrodrizzle} reductions of the 11 broad- and medium-bands published in \cite{shar22} and five additional narrow-bands taken from GO-15949. The narrowband data are processed using the same implementation of the \texttt{Astrodrizzle} pipeline as the other HST data from \cite{shar22} and will be described in more detail in J. Rigby et al. (2023, in prep.). All imaging data was drizzled to produce final data products with a common pixel scale of 0.03$''$. Since narrowband filters sample the continuum plus emission line flux, we also employ the associated, adjacent continuum filters (``Continuum'') to subtract off the continuum flux density for each line.
The HST photometry used in this study appears in Table \ref{tab1}.

\begin{deluxetable*}{ccccccccccc}




\tablecaption{HST broadband and narrowband photometry used in this study (Section \ref{sec:Section2_1_HST_broad_narrow_imaging}).}

\tablenum{1}

\tablehead{\colhead{Instrument/Mode} & \colhead{Filter\tablenotemark{a}} & \colhead{Exposure time (s)} & \colhead{Rest-frame features\tablenotemark{b,c}} 
} 
\startdata
WFC3/UVIS & F390W & 5853 & Ly$\alpha$ \textbf{Continuum} \\
WFC3/UVIS & F410M & 13,285 & Ly$\alpha$ \\
WFC3/UVIS & F555W & 5616 & $\beta$ continuum ($\lambda_{\rm eff} \simeq 1600$ \AA) \\
WFC3/UVIS & F606W & 7878 & $\beta$ continuum ($\lambda_{\rm eff} \simeq 1800$ \AA) \\
ACS/WFC & F814W & 5280 & $\beta$ continuum ($\lambda_{\rm eff} \simeq 2400$ \AA) \\
WFC3/IR & F126N & 11212 &  \OII\ $\lambda$3726,3729 \AA \\
WFC3/IR & F128N & 11212 & \OII\ \textbf{Continuum} \\
WFC3/IR & F153M & 5612 & \Hb\ \& \OIII\ \textbf{Continuum} \\
WFC3/IR & F164N & 5612 & H$\beta$ \\
WFC3/IR & F167N & 5612 & \OIII\ $\lambda$4959 \AA \\
\hline
\hline
\enddata

\tablenotetext{a}{In the naming convention for HST filters, an initial following the filter number denotes: W;broad, M;medium, and N;narrow}
\tablenotetext{b}{At the redshift of the Sunburst Arc ($z=2.37$).}
\tablenotetext{c}{``Continuum'' indicates the associated, adjacent filter used for continuum subtraction of each line (Section \ref{sec:Sec_2_5_emission line measurements}).}


\label{tab1}
\end{deluxetable*}

\subsection{Resolving LyC Leaking and Non-leaking Regions on small ($< 100$ pc) scales}
\label{sec:Section2_2_identification of LyC and non leaking regions}

The powerful combination of HST's clear imaging with a lensing magnification on the Sunburst Arc provides an effective spatial resolution down to tens of parsecs \citep{rive19,vanz22, shar22,dieg22}. The HST imaging of the Sunburst Arc reveals that \textit{only} one compact star-forming region emits LyC radiation while other parts of the galaxy do not \citep{rive19}. 
The lensed galaxy images appear as four individual arcs in the sky. Across the four lensed arcs, a single leaking region is multiply lensed, resulting in 12 detectable clumps with different magnifications and lines of sight of the same physical region. Like the leaking region, the non-leaking regions of the galaxy are multiply lensed; confirmed multiple images of individual regions are described in the strong lens model \citep[see Table 2 in][]{shar22}.
Images of the Sunburst Arc are shown in Figure \ref{fig0}. 
We leverage the high spatial resolution of HST to isolate the emission from the individual regions, and characterize the spatially resolved physical conditions of leaking and non-leaking regions. 

\subsection{UV-continuum Slope ($\beta$) Measurements}
\label{subsubsec:Sec_2_3_uvslope measurements}
We measure the UV-continuum slope $\beta$ of the multiple lensed images of the LyC leaking region, as well as images of the non-leaking regions within the galaxy. We measure $\beta$ from three HST broadband images---WFC3/UVIS F555W and F606W, and ACS/WFC F814W---that cover the rest-frame UV-continuum ($\sim$1600 --- 2400 $\angstrom$) of the galaxy at $z$ = 2.37. Prior to measuring $\beta$, we correct the three bands for Milky Way reddening\footnote{NASA/IPAC Galactic Dust Reddening and Extinction tool: \url{https://irsa.ipac.caltech.edu/applications/DUST/}} ($E(B-V) = 0.094$) by adopting the \cite{card89} reddening law with $R_{V} = 3.1$.
The images are then PSF-matched to the longest wavelength data available (accounting for different spatial resolution across the available HST data, which is a 
 FWHM $= 0\farcs 15$ for the reddest narrowband data).

We perform aperture photometry for individual clump images using circular apertures with a diameter of 8 pixels ($=0\farcs 24$), which captures most of the emission from the largely unresolved clumps while avoiding contamination from other neighboring structures. 
The $\beta$ is then derived by fitting the measured fluxes with the associated wavelengths for each region following the relationship \citep{calz94}:
\begin{eqnarray}
f_{\lambda} \propto {\lambda}^{\beta}
\label{Eq1}
\end{eqnarray}
where $f_{\lambda}$ is the flux density per unit wavelength ($\rm{erg} \ \rm{s^{-1}} \ \rm{cm^{-2}} \ \angstrom^{-1}$) and $\lambda$ is the effective wavelength for each of F555W, F606W, and F814W (5308\AA, 5887\AA, and 8045\AA, respectively). We also measure the integrated galaxy-wide UV slope by stacking pixels of the West Arc, which is a complete image of the whole galaxy \citep[i.e., Arc 3 in Figure \ref{fig0}, see also][]{rive19,shar22}. The measured UV slopes are reported in Table \ref{tab2} and span a range of $\beta$ values from $\simeq-2.9$ to $\simeq-2.2$.

\subsection{Reddening Corrections}
\label{sec:Sec2_4_emission line measurements}
We compute nebular reddening corrections for the narrowband imaging data (i.e., emission line images) using the \cite{calz00} law and the average $E(B-V)_{\rm gas} = 0.195 \pm 0.025$ measured by \cite{main22} from the H$\alpha$/H$\beta$ Balmer decrement in moderate resolution, rest-frame optical spectra of the Sunburst Arc obtained with the Folded-port InfraRed Echellette (FIRE; \citealt{simc13}) spectrometer mounted on the Magellan-I Baade Telescope. The adoption of a uniform reddening correction is consistent with the conclusions of \cite{main22}; they note that there is no empirical evidence for large differences in the Balmer decrements of ground-based spectra targeting leaking vs. non-leaking regions. The assumption of a uniform Balmer reddening correction is also consistent with the stellar reddening ($E(B-V)_{\rm stellar}$) derived from the FUV SED modelling \citep{chis19}, which finds statistically consistent values for both the leaking and non-leaking regions (i.e., $E(B-V)_{\rm stellar}$ of $0.08 \pm 0.02$ vs. $0.06 \pm 0.01$ \citep[Table 3 in][]{main22}). Interestingly, the approximate ratio of $E(B-V)_{\rm stellar}$ to $E(B-V)_{\rm gas}$ in the Sunburst Arc is also remarkably consistent with the canonical ratio ($\sim0.4$) measured for local starburst galaxies \citep{calz00}.

It is important to note that the stellar and nebular reddening terms 
are estimated from ground-based observations (optical and NIR slit 
spectroscopy, respectively), which include emission that is averaged 
over larger angular scales (i.e., ground-based seeing of 
$\sim0.6-0.8''$) than the HST imaging that we use to measure 
spatially resolved UV slopes. This means that the true local reddening 
terms affecting spatially resolved regions within the Sunburst Arc could vary across the individual star-forming knots, and partially contribute to the different UV slopes measured in leaking vs. non-leaking regions. 
However, given the overall low reddening measured for the Sunburst Arc, 
spatially-variable dust reddening alone cannot account for the UV slope 
differences measured for leaking and non-leaking regions. Assuming the 
maximum possible stellar reddening difference in which the isolated 
LyC-leaking regions are totally dust-free ($E(B-V)_{\rm stellar} = 0$) and the non-leaking regions have $E(B-V)_{\rm stellar} = 0.08$, a 
\cite{redd16} extinction law can only produce a difference in the observed UV slope of $\Delta \beta \simeq 0.3$ \citep[][see also their Figure 5]{chis22}. This difference is too small to fully explain the different UV slopes measured for the leaking vs. non-leaking regions in the Sunburst Arc (i.e., $-2.9 \pm 0.1$ vs. $-2.2 \pm 0.2$ (Section \ref{sec:Sec_3_results})).

As we will discuss further the spatially-resolved UV slope properties in Sections \ref{sec:Sec4_1_LyC_extm_properties} and \ref{sec:Section4_2_UV_slope_Pure_stellar}, in summary, there is extremely low internal reddening across the entirety of the Sunburst Arc, and no evidence for significant spatial variations. The uniformly small reddening affecting different lines of sight toward the Sunburst Arc indicates that differences in the stellar population ages and ISM ionization fractions are primarily responsible for the different UV slope values that we measure in spatially resolved regions within the Sunburst Arc.

\subsection{Emission Line Measurements: {\rm{\lya, \Hb, \OII}}, and \rm{\OIII}}
\label{sec:Sec_2_5_emission line measurements}
Using the HST narrowband imaging described in Section \ref{sec:Section2_1_HST_broad_narrow_imaging}, we measure the optical emission line ratios (\OIII /\OII\ and \OIII /\Hb) and the \lya\ escape fraction (\Lyaesc ) for investigating nebular ionization state and local \lya\ escape processes, respectively, of the leaking region and the non-leaking regions.
Measuring emission line flux requires first subtracting off the underlying continuum emission. 
We subtract the continuum by fitting the shape of the spectral energy distribution (SED) of the continuum flux density, $f_{\lambda}$, to all of the available HST data$-$except for the strongest emission line filters (i.e., F126N, F164N, and F167N as in Table \ref{tab1})$-$of the four multiple images of the whole galaxy contained in each of Arcs 1 and 2 (Figure \ref{fig0}) pixel-wise.
We feed this measured SED into the \texttt{stsynphot} and \texttt{synphot} packages \citep{lim19}, to compute the count rates contributed by the continuum to each filter containing an emission line, as well as its corresponding continuum filter.

Specifically, we compute the count rate corresponding to a source whose spectrum is a delta-function emission line of known flux ($10^{-15} \rm{erg \ s^{-1} \ cm^{-2}}$).
For each emission line, this results in a scaling factor to translate the measured count rate to line flux.
Then, for each narrowband filter containing an emission line, we use these predicted count rates to scale the continuum filter, and we then subtract that scaled filter image from the narrow band image of the emission line.

The continuum SEDs and the associated narrowband filters are shown in Figure \ref{fig8}.
The 1$\sigma$ fractional uncertainties in the estimated continuum  are 10\%, 6\%, and 6\%, respectively, for \Lya , \OII , and both \Hb\ and \OIII; these uncertainties include both statistical and systematic sources of error in the SED fit and the continuum normalization using adjacent filters. A more detailed description of the continuum fitting and subtraction procedure will appear in a forthcoming paper (J. Rigby et al., in prep.)

We then apply Milky Way reddening corrections to the continuum-subtracted narrowband images. We also apply PSF-matching to the images, with the exception of the F164N and F167N filters (\Hb\ and \OIII\, respectively), both of which are natively at the final, convolved PSF (FWHM of 0.15 \arcsec). Following the same aperture photometry procedure as the UV slope measurement described above (Section \ref{subsubsec:Sec_2_3_uvslope measurements}), we measure the key emission line fluxes (i.e., \Lya, \Hb, \OII, and \OIII ) of the leaking region, the non-leaking regions, and the entire galaxy using the continuum-subtracted narrowband flux densities and the bandwidths of each filter. We also correct all emission lines, except for \Lya, for internal nebular reddening using the same Balmer decrement as in Section \ref{sec:Sec2_4_emission line measurements}. Internal reddening corrections for \Lya\ are non-trivial due to significant resonant scattering at its line center ($\tau_{\rm 0} \gg 1$), such that it is virtually impossible to correct for nebular reddening appropriately \citep[e.g.,][]{neuf91,verh15,dijk19}). We thus do not correct \Lya\ emission for internal reddening effect. Ultimately we do not use the \Lya\ images to measure any line ratios that require de-reddened \Lya\ emission.

\subsection{Emission Line Ratios and \Lya\ escape fraction (\Lyaesc)}
\label{sec:Sec_2_6_emission line ratio_LyA_escape}
We use the emission line fluxes measured from the narrowband filters to measure several standard nebular emission line ratios, including the \OIII /\OII\ line ratio, given by 
\begin{eqnarray}
\OIII /\OII\ = \frac{\OIII\ 5007 \angstrom}{\OII\ 3727,3729 \angstrom}
\label{Eq2}
\end{eqnarray}
and the \OIII /\Hb\ ratio, given by 
\begin{eqnarray}
\OIII /H\beta = \frac{\OIII\ 5007 \angstrom}{H\beta~4863 \angstrom}
\label{Eq3}
\end{eqnarray}
We use the \OIII $\lambda$5007 flux throughout the paper for a direct comparison of our results with the literature. The \OIII $\lambda$5007 is derived from the measured \OIII $\lambda$4959 flux by multiplying the line ratio by 2.98, which is fixed by atomic physics \citep{stor00}.

We also measure the \Lya\ escape fraction (\Lyaesc) from the ratio of the observed flux of \Lya\ to the expected flux of \Lya\ based on the observed \Hb\ line flux. 
In practice \Lyaesc\ is simply measured as the ratio of \Lya\ to \Hb\ multiplied by a scalar that comes from assuming case B recombination, which means that the \textit{intrinsic} \Lya\ flux is estimated from the Hydrogen recombination physics.
Specifically, we calculate the intrinsic \Lya\ flux by multiplying the \Hb\ flux by 23.3, which is the appropriate scaling factor for Case B recombination for an electron temperature $T_{\rm e} = 10000$ K and an electron number density $n_{\rm e} = 100 \ \rm{cm^{-3}}$ \citep{dopi03}.
While the definition of \Lyaesc\ adopted in this paper is consistent with the literature, it should be noted that our \Lyaesc\ is measured on scales of \textit{resolved} star-forming regions within a galaxy \citep[e.g.,][]{woff13,rive15}. 
This differs from the \Lyaesc\ measured based off the entire galaxy in the literature \cite[e.g., ][]{henr15,yang17}.
Thus, our \Lyaesc\ measured from the resolved star-forming regions should be considered the ``local'' line of sight observed \Lya\ escape fraction rather than the global galaxy property.

\begin{figure*}
\includegraphics[width=\linewidth]{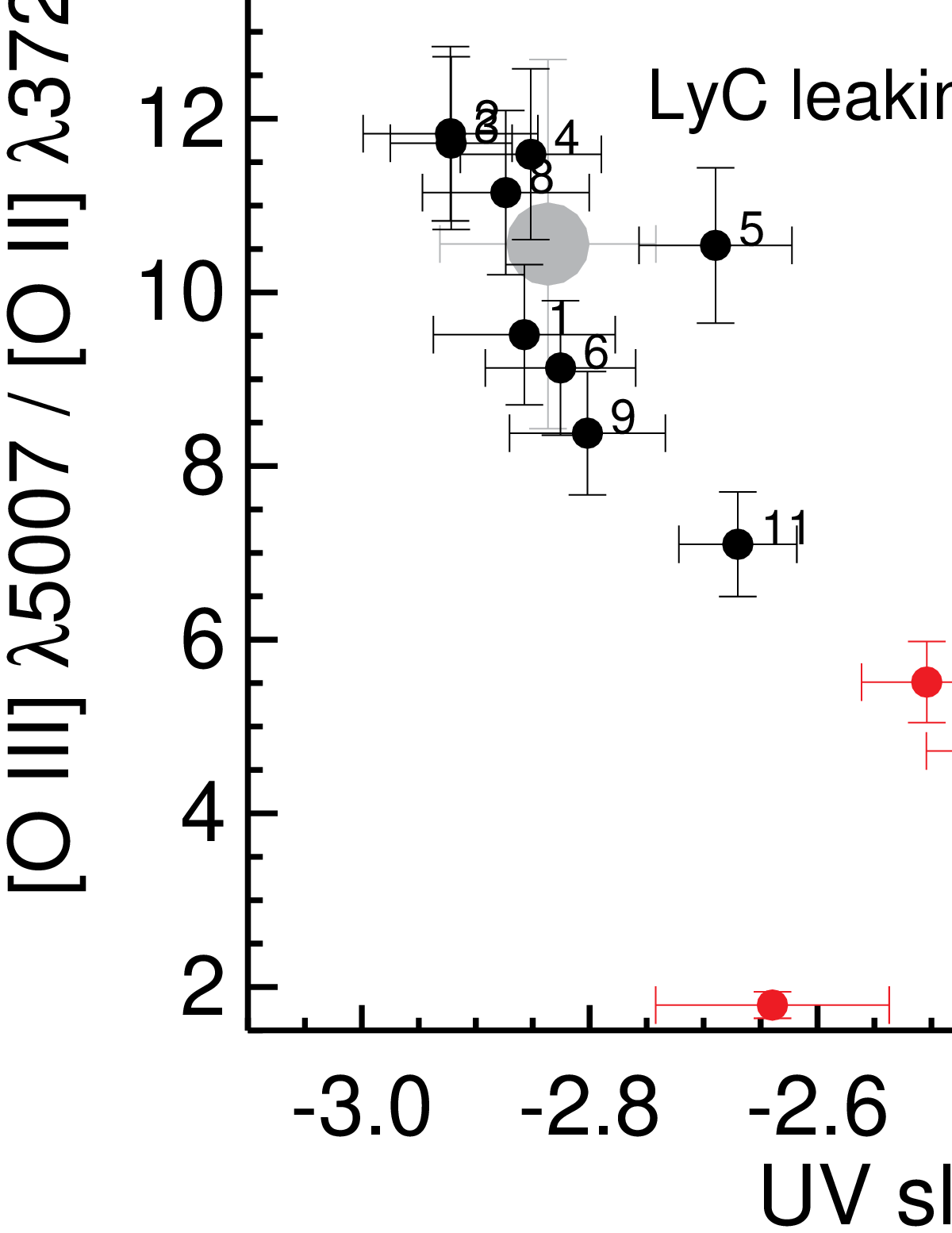}
\caption{\textbf{Left:} \OIII /\OII\ vs. UV-continuum slope ($\beta$) for the leaking region (black), the non-leaking regions (red), and the entire galaxy (blue) as marked in Figure \ref{fig0}. The light grey and the pink circles with error bars indicate the mean value and the 1$\sigma$ population deviation of each region. \textbf{Right:} Same as the left panel, but for \OIII/\Hb.
In the Sunburst Arc, the multiple-lensed images of the LyC leaking region (black) show the remarkably blue UV-continuum slope $\beta \simeq -2.9$ and high ionization state of surrounding ISM (\OIII/\OII\ = $11 \pm 3$ and \OIII/\Hb\ = $6.8 \pm 0.4$), which is distinctly extreme compared to the non-leaking regions (red) and the entire galaxy (blue).
The leaking region's extremely blue $\beta$ and high ionization-sensitive line ratios (corresponding to ionization parameter of ${\rm{log}}U$ of $\simeq -2$, Section \ref{sec:Sec4_1_LyC_extm_properties}) suggest that LyC photons are produced and escape from a local, highly ionized compact region with extreme stellar populations (Section \ref{sec:Sec4_discussion}).
\label{fig1}}
\end{figure*}

\subsection{\Hb\ Flux Calibration}
\label{sec:Sec_2_7_Hbeta flux calibration}
Comparing our measurements to other observations of the Sunburst Arc, including the ground-based spectroscopy presented in \cite{main22}, as well as unpublished HST WFC3/IR G141 grism spectroscopy (PID: 15101) (J. Rigby et al., in prep.), we find that the \Hb\ flux measured from the F164N WFC3/IR filter is systematically $\sim$60\% lower than other measurements of the same emission line in spectra taken with both Magellan/FIRE and the WFC3/IR G141 grism (J. Rigby, in prep.). This offset is much larger than the statistical and systematic uncertainties in the measured F164N \Hb\ flux, and cannot be explained by any treatment of the continuum subtraction (i.e., a large negative continuum flux density, which is non-physical, would be required to bring F164N into agreement with the WFC3/IR grism). The large difference between narrowband and grism \Hb\ fluxes is especially puzzling considering that F164N and the G141 grism are part of the same WFC3/IR instrument and calibration pipeline. After a thorough exploration of all possible systematic, reduction, and analysis effects, we conclude that the calibration data available in the WFC3/IR pipeline for F164N---a scarcely used narrowband filter--- is likely out of date, and that the filter throughput either was not correctly calibrated, or that it has substantially degraded since the most recent calibration observations were taken.
Therefore, we ad-hoc increase all \Hb\ flux measurements by 60$\%$ to account for this uncertainty.

It is important to note that this correction does not qualitatively change our comparison of the emission line properties of the leaking vs. non-leaking regions, because it is applied uniformly to the entire F164N image, and therefore shifts the measured line ratio of \OIII /\Hb\ and \Lyaesc\ and equivalent width of \Hb\ (EW (\Hb )) in the same direction. 

\subsection{Emission Line Equivalent Widths}
\label{sec:Sec_2_8_Equivalent Width Measurements}
The emission line fluxes and underlying continuum flux density measurements described above can also be used to measure the equivalent width (EW) of the nebular emission lines. We compute the observed-frame EW$_{\rm{obs}}$ of each emission line by dividing the attenuation uncorrected line flux by the underlying continuum flux density (assumed to be constant), and then compute the rest-frame equivalent width as EW$_{\rm rest}$ $=$ EW$_{\rm obs}$/($1+z)$. The resulting emission line EW values are useful for contextualizing the EW properties of the clumps (i.e., the leaking region and the non-leaking regions) we observe in the Sunburst Arc with other extreme star-forming galaxies in the literature.

\section{Results} 
\label{sec:Sec_3_results}
Our analysis pins down the properties of the LyC leaking region on small scales $< 100$ pc within a galaxy. Specifically, we compare the relationships between the UV slope $\beta$, the ionization sensitive line ratios \OIII /\OII\ and \OIII /\Hb , and \Lya\ escape fraction \Lyaesc\ for individual resolved regions within the Sunburst Arc, as well as for the entire galaxy.
Due to contamination from a foreground galaxy associated with an intervening absorption system \citep{lope20} and a foreground star, the images 7 and 12 of the leaking region are excluded in this analysis.
All measurements are reported in Table \ref{tab2}.

It is noteworthy that all of the presented parameters are lensing magnification \textit{independent} as they are intrinsically the \textit{ratio} of measured parameters, which means that the lensing magnification cancels out.
Rather, what the lensing magnification uniquely provides in our analysis is spatially-resolved morphology of the Sunburst Arc on small scales sufficient to isolate the leaking region from the non-leaking regions \citep[see][for further details about the lens model]{shar22}.

\begin{figure*}
\includegraphics[width=\linewidth]{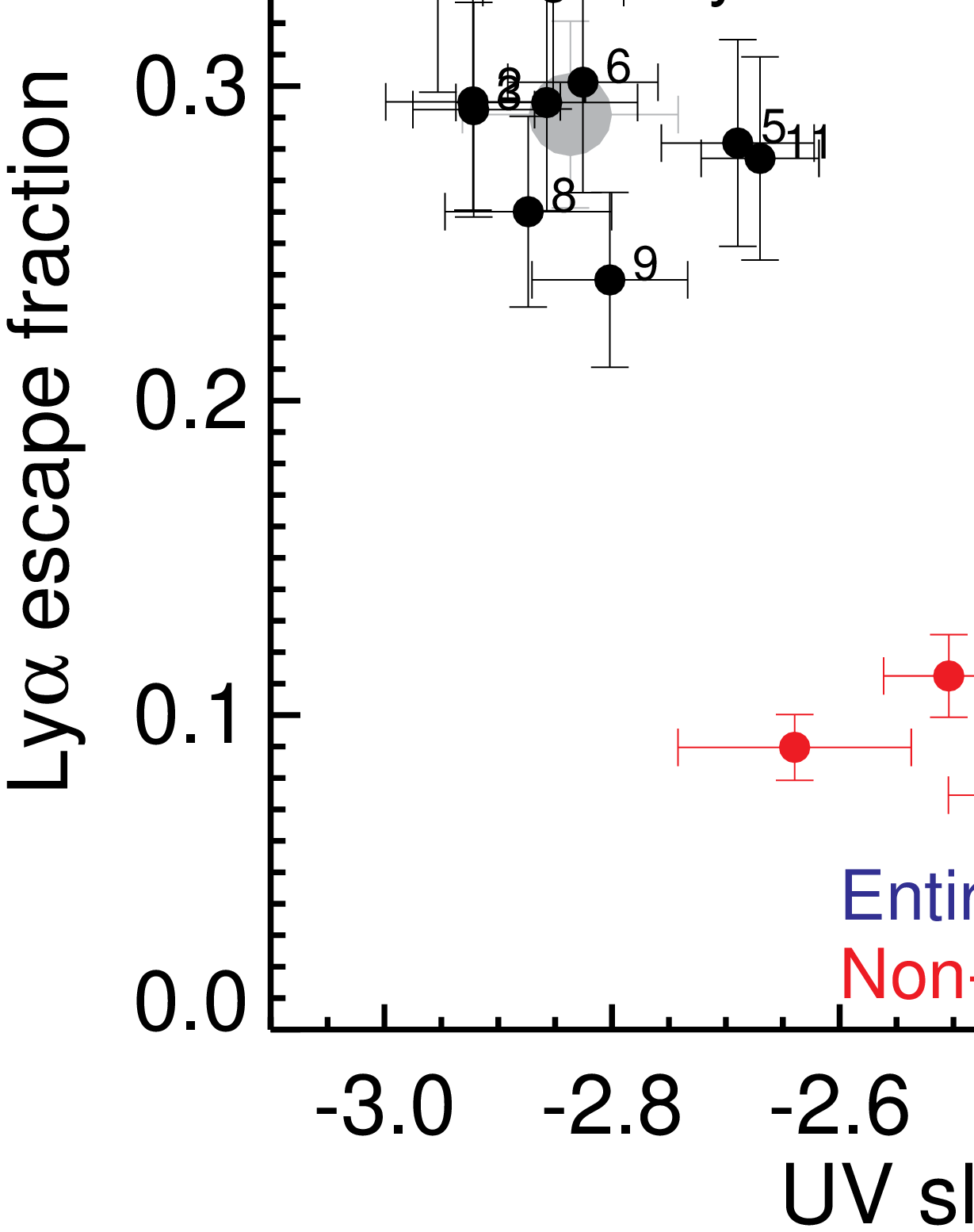}
\caption{Ly$\alpha$ escape fraction (\Lyaesc ) vs. $\beta$ (left) and  \OIII /\OII\ vs. \Lyaesc\ (right) relations for the leaking region, the non-leaking regions, and the entire galaxy. The format is the same as in Figure \ref{fig1}. 
Combined with the very blue $\beta$ and high \OIII /\OII\ ratio, the leaking region is clearly characterized by relatively high \Lyaesc , the presence of extreme stellar populations, and highly ionized nebular state (Sections \ref{sec:Sec4_1_LyC_extm_properties} and \ref{sec:Section4_2_UV_slope_Pure_stellar}).
It also implies the physical connection between LyC and \lya\ escape processes on sub-kpc scales such that the LyC leaking region features a higher escape fraction of \Lya\ as well, likely due to the preferred low $N$(\HI ) channels for both photons. (Section \ref{sec:Section4_3_LyC_LyA}).
\label{fig2}}
\end{figure*}

\subsection{Very Blue UV Slope and High Ionization State of the LyC-leaking Region}
\label{subsec:Sec_3_1_UV slope vs. O32}
This analysis constrains, for the first time at any redshift, the 
UV slope $\beta$, as well as the \OIII /\OII\ \& \OIII /\Hb\ line 
ratios, of a spatially resolved leaking star-forming region within 
a galaxy (rather than galaxy-integrated). The individual images of the 
leaking region show distinctly blue UV-continuum slope and high 
ionization compared to any other non-leaking regions within the galaxy. This trend 
is shown in Figure \ref{fig1}, where it is clear that 
the multiply lensed leaking region (the numbered black points in the 
figure) occupies the parameter space of very blue UV slope ($\beta =
-2.9 \pm 0.1$) and high ionization (\OIII /\OII\ = $11 \pm 3$, and 
\OIII /\Hb\ = $6.8 \pm 0.4$). 
This is in stark contrast to the non-leaking regions (i.e., the red points) 
which show systematically redder UV slope ($\beta \simeq -2.2$) and 
lower ionization state (\OIII /\OII $\simeq 4$, \OIII /\Hb\ $\simeq 4$).

These extreme values of $\beta$ and \OIII /\OII\ of the leaking region significantly ($\gtrsim 7 \sigma$) deviate from the non-leaking regions which have $\beta$ and \OIII /\OII\ of $-2.2 \pm 0.2$ and $4 \pm 0.6$, respectively. Also, a comparison with the whole galaxy-integrated $\beta$ of $-2.2 \pm 0.03$ and \OIII /\OII\ of $4 \pm 0.3$ (i.e., represented by the blue diamond in Figure \ref{fig1}) shows that the UV slope and ionization state of the leaking region are indeed extreme relative to the galaxy as a whole.

\subsection{Blue UV Slope versus High \Lya\ Escape From the LyC-leaking Region}
\label{subsec:Sec_3_2UV slope vs. LyA escape}
Plotting $\beta$ and \Lyaesc\ for all measured regions within the Sunburst Arc (i.e., left panel of Figure \ref{fig2}) reveals a consistent separation between the LyC leaking region and the other regions of the galaxy. The blue $\beta$ and high \Lyaesc\ of the leaking region are clearly distinct from the redder $\beta$ and lower \Lyaesc\ of the non-leaking regions, and there is a broad anti-correlation between $\beta$ and \Lyaesc. We measure typical \Lyaesc\ for the leaker and non-leakers of $0.3 \pm 0.03$ and $0.13 \pm 0.07$, respectively. Also similar to what we saw with $\beta$ and \OIII /\OII\ and \OIII/\Hb\ ratios, the galaxy-integrated \Lyaesc\ ($0.14 \pm 0.02$) matches the non-leaker values. 

Although the absolute values of \Lyaesc\ in this paper are subject to a systematic uncertainty resulting from the problematic \Hb\ narrowband imaging flux calibration (Section \ref{sec:Sec_2_7_Hbeta flux calibration}), we note that the trend of the higher \Lyaesc\ of the leaking region than non-leaking regions (i.e., Figure \ref{fig2}) does not change because the flux offset applies to all regions of the galaxy.

\subsection{High \Lya\ Escape Fraction versus High Ionization State for the LyC-leaking Region}
\label{subsec:Sec_3_3_LyA escape and high ionization}

The relationships between \Lyaesc\ and ionization state (\OIII /\OII ) within the Sunburst Arc reveal a consistent picture as the preceding sections. 
The right panel of Figure \ref{fig2} shows that the leaking region occupies the parameter space of high \Lyaesc\ and high \OIII /\OII\ while the non-leaking regions have low \Lyaesc\ and low \OIII /\OII. 
The galaxy-integrated values are consistent with those of the non-leaking regions. 
This trend of high \Lya\ escape fraction and high ionization of the leaking region condenses the relations seen in the left panels of Figures \ref{fig1} and \ref{fig2} showing that the UV slope tracks closely with both the ionization state (Figure \ref{fig1}) and \Lyaesc\ (Figure \ref{fig2}) by directly relating the ionization state with \Lyaesc .

This result makes physical sense, with the right panel of Figure \ref{fig2} clearly showing that the escape of \Lya\ is typically related to the nebular ionization state on the scale of star-forming clumps, such that \Lya\ photons locally escape more easily from the highly ionized star-forming regions than the low ionization regions. We also note that the galaxy-integrated \Lyaesc\ and \OIII /\OII\ look like the resolved non-leaking regions, all of which have \Lyaesc\ and ionization properties that are broadly consistent with those observed in low redshift Green Pea galaxies \citep[e.g.,][]{yang17,flur22}. Additionally, while the interpretation on the relation between the escape of \Lya\ photons and the nebular ionization state makes physical sense overall, it should also be noted that the actual physical correlations may be more complex, in the sense that a high \Lyaesc\ region does not always correspond to a high \OIII /\OII\ region \cite[e.g., Haro 11,][]{keen17,ostl21}.

\subsection{High Equivalent Width Emission Lines of the LyC-leaking Region}
\label{subsec:Sec3_4_High EW with table}
LyC leakers are often characterized by the high equivalent width (EW) of emission lines (notably, \Lya , \Hb , \OIII $\lambda$5007, and \OII$\lambda$3727,3729) \citep[e.g.,][]{izot16,izot18}. 
We see the same qualitative relationship between emission line EW and LyC leakage, notably on \textit{sub-galactic scales} in the emission line EWs of the leaking region, non-leaking regions, and the entire Sunburst Arc galaxy (Table \ref{tab2}).
The median EWs of the leaking region are typically a factor of 2$-$5 (depending on which specific line is compared) higher than those of the non-leaking regions.
The EW values measured for the entire, integrated galaxy emission are similar to those of the non-leaking region.
This is to be expected because the leaking region is only a very small part of the galaxy ($\sim 17 \%$ of the UV-continuum light) while the non-leaking regions dominate the galaxy.

Consistent with previous results about bluer UV slope, higher ionization state, and higher \lya\ escape fraction (Sections \ref{subsec:Sec_3_1_UV slope vs. O32}, \ref{subsec:Sec_3_2UV slope vs. LyA escape}, and \ref{subsec:Sec_3_3_LyA escape and high ionization}), the leaking region’s higher emission line EWs suggest that ionizing photons escape from a specific star-forming region with extreme physical properties inside the galaxy.

\begin{deluxetable*}{ccccccccccc}




\tablecaption{The measured properties across the Sunburst galaxy, showing the distinctly extreme properties of the LyC leaking region compared to the non-leaking regions and the entire galaxy (Section \ref{sec:Sec_3_results}).}

\tablenum{2}

\tablehead{\colhead{Region and ID} & \colhead{UV slope ($\beta$)} & \colhead{\OIII/\OII} & \colhead{\OIII/H$\beta$} & \colhead{$f_{\rm esc}^{\rm Ly\alpha}$\tablenotemark{a}} & \colhead{EW(Ly$\alpha$)\tablenotemark{b}} & \colhead{EW(H$\beta$)\tablenotemark{b}} & \colhead{EW(\OIII)\tablenotemark{b}} & \colhead{EW(\OII)\tablenotemark{b}} \\
\colhead{} & \colhead{1600$-$2400 $\angstrom$} & \colhead{} & \colhead{} & \colhead{} & \colhead{[$\angstrom$]} & \colhead{[$\angstrom$]} & \colhead{[$\angstrom$]} & \colhead{[$\angstrom$]} 
} 
\startdata
LyC leaking\tablenotemark{c} &  &  &  &  &  &  &  & \\
\hline
 1 & $-2.86 \pm 0.08$ & $9.51 \pm 0.81$ & $6.77 \pm 0.57$ & $0.29 \pm 0.03$ & $48.5 \pm 4.8$ & $163.9 \pm 9.8$ & $1147.2 \pm 68.8$ & $47.4 \pm 2.8$ \\
 2 & $-2.92 \pm 0.08$ & $11.83 \pm 1.00$ & $6.89 \pm 0.59$ & $0.29 \pm 0.03$ & $43.1 \pm 4.3$ & $161.4 \pm 9.7$ & $1152.5 \pm 69.1$ & $35.7 \pm 2.1$ \\
 3 & $-2.92 \pm 0.05$ & $11.72 \pm 0.99$ & $6.57 \pm 0.56$ & $0.29 \pm 0.03$ & $39.9 \pm 4.0$ & $158.8 \pm 9.5$ & $1080.5 \pm 64.8$ & $35.1 \pm 2.1$ \\
 4 & $-2.85 \pm 0.06$ & $11.59 \pm 0.98$ & $6.77 \pm 0.57$ & $0.33 \pm 0.04$ & $42.9 \pm 4.3$ & $150.5 \pm 9.0$ & $1054.9 \pm 63.3$ & $35.3 \pm 2.1$ \\
 5 & $-2.69 \pm 0.07$ & $10.54 \pm 0.89$ & $6.75 \pm 0.57$ & $0.28 \pm 0.03$ & $42.1 \pm 4.2$ & $126.8 \pm 7.6$ & $887.1 \pm 53.2$ & $34.7 \pm 2.1$ \\
 6 & $-2.83 \pm 0.07$ & $9.13 \pm 0.77$ & $6.89 \pm 0.58$ & $0.30 \pm 0.04$ & $44.0 \pm 4.4$ & $156.8 \pm 9.4$ & $1116.2 \pm 67.0$ & $47.3 \pm 2.8$ \\
 8 & $-2.87 \pm 0.07$ & $11.15 \pm 0.95$ & $6.14 \pm 0.52$ & $0.26 \pm 0.03$ & $40.5 \pm 4.0$ & $178.9 \pm 10.7$ & $1135.0 \pm 68.1$ & $38.6 \pm 2.3$ \\
 9 & $-2.80 \pm 0.07$ & $8.38 \pm 0.71$ & $6.01 \pm 0.51$ & $0.24 \pm 0.03$ & $41.6 \pm 4.2$ & $176.7 \pm 10.6$ & $1095.0 \pm 65.7$ & $51.2 \pm 3.1$ \\
 10 & $-2.95 \pm 0.08$ & $14.62 \pm 1.24$ & $6.29 \pm 0.53$ & $0.34 \pm 0.04$ & $32.9 \pm 3.3$ & $128.7 \pm 7.7$ & $838.4 \pm 50.3$ & $21.6 \pm 1.3$ \\
 11 & $-2.67 \pm 0.05$ & $7.10 \pm 0.60$ & $5.83 \pm 0.50$ & $0.28 \pm 0.03$ & $44.2 \pm 4.4$ & $138.3 \pm 8.3$ & $830.9 \pm 49.9$ & $46.3 \pm 2.8$ \\
\hline
\textbf{Median\tablenotemark{d}} & $-2.9 \pm 0.03$ & $10.6 \pm 0.7$ & $6.8 \pm 0.1$ & $0.3 \pm 0.01$ & $43 \pm 1.3$ & $159 \pm 5.7$ & $1095 \pm 41$ & $39 \pm 2.8$ \\
\hline
\hline
Non-leaking\tablenotemark{e} &  &  &  &  &  &  &  & \\
\hline
 1 & $-2.16 \pm 0.03$ & $3.13 \pm 0.27$ & $4.84 \pm 0.41$ & $0.21 \pm 0.03$ & $17.5 \pm 1.8$ & $36.3 \pm 2.2$ & $180.7 \pm 10.8$ & $26.9 \pm 1.6$ \\
 2 & $-2.32 \pm 0.02$ & $3.11 \pm 0.26$ & $3.47 \pm 0.29$ & $0.22 \pm 0.03$ & $13.0 \pm 1.3$ & $34.2 \pm 2.1$ & $122.4 \pm 7.3$ & $18.5 \pm 1.1$ \\
 3 & $-2.36 \pm 0.04$ & $3.17 \pm 0.27$ & $3.50 \pm 0.30$ & $0.19 \pm 0.02$ & $9.7 \pm 1.0$ & $28.7 \pm 1.7$ & $104.0 \pm 6.2$ & $15.1 \pm 0.9$ \\
 4 & $-2.19 \pm 0.00$ & $3.50 \pm 0.30$ & $4.07 \pm 0.35$ & $0.17 \pm 0.02$ & $10.0 \pm 1.0$ & $31.4 \pm 1.9$ & $131.8 \pm 7.9$ & $18.1 \pm 1.1$ \\
 5 & $-2.22 \pm 0.02$ & $3.96 \pm 0.34$ & $4.72 \pm 0.40$ & $0.16 \pm 0.02$ & $11.2 \pm 1.1$ & $35.5 \pm 2.1$ & $173.2 \pm 10.4$ & $21.0 \pm 1.3$ \\
 6 & $-2.46 \pm 0.05$ & $4.72 \pm 0.40$ & $4.23 \pm 0.36$ & $0.07 \pm 0.01$ & $4.3 \pm 0.4$ & $37.8 \pm 2.3$ & $165.0 \pm 9.9$ & $15.2 \pm 0.9$ \\
 7 & $-2.15 \pm 0.00$ & $3.34 \pm 0.28$ & $4.44 \pm 0.38$ & $0.10 \pm 0.01$ & $9.3 \pm 0.9$ & $52.6 \pm 3.2$ & $240.1 \pm 14.4$ & $33.7 \pm 2.0$ \\
 8 & $-2.25 \pm 0.01$ & $3.51 \pm 0.30$ & $4.25 \pm 0.36$ & $0.11 \pm 0.01$ & $11.5 \pm 1.1$ & $52.7 \pm 3.2$ & $230.6 \pm 13.8$ & $30.3 \pm 1.8$ \\
 9 & $-2.39 \pm 0.00$ & $5.21 \pm 0.44$ & $4.71 \pm 0.40$ & $0.13 \pm 0.02$ & $7.2 \pm 0.7$ & $39.6 \pm 2.4$ & $192.4 \pm 11.5$ & $15.3 \pm 0.9$ \\
 10 & $-2.33 \pm 0.03$ & $4.46 \pm 0.38$ & $4.21 \pm 0.36$ & $0.22 \pm 0.03$ & $14.8 \pm 1.5$ & $38.0 \pm 2.3$ & $165.6 \pm 9.9$ & $17.2 \pm 1.0$ \\
 11 & $-1.89 \pm 0.00$ & $3.45 \pm 0.29$ & $5.61 \pm 0.48$ & $0.25 \pm 0.03$ & $63.4 \pm 6.3$ & $108.7 \pm 6.5$ & $623.2 \pm 37.4$ & $77.7 \pm 4.7$ \\
 12 & $-2.03 \pm 0.03$ & $3.26 \pm 0.28$ & $3.91 \pm 0.33$ & $0.18 \pm 0.02$ & $46.1 \pm 4.6$ & $91.1 \pm 5.5$ & $364.9 \pm 21.9$ & $52.7 \pm 3.2$ \\
 13 & $-1.96 \pm 0.00$ & $3.55 \pm 0.30$ & $4.82 \pm 0.41$ & $0.25 \pm 0.03$ & $59.8 \pm 6.0$ & $90.9 \pm 5.5$ & $448.0 \pm 26.9$ & $56.5 \pm 3.4$ \\
 14 & $-1.88 \pm 0.05$ & $3.37 \pm 0.29$ & $5.32 \pm 0.45$ & $0.24 \pm 0.03$ & $88.0 \pm 8.8$ & $152.6 \pm 9.2$ & $826.8 \pm 49.6$ & $105.4 \pm 6.3$ \\
 15 & $-2.03 \pm 0.01$ & $2.86 \pm 0.24$ & $4.57 \pm 0.39$ & $0.04 \pm 0.00$ & $13.0 \pm 1.3$ & $129.5 \pm 7.8$ & $597.3 \pm 35.8$ & $112.0 \pm 6.7$ \\
 16 & $-2.12 \pm 0.02$ & $2.94 \pm 0.25$ & $3.82 \pm 0.32$ & $0.04 \pm 0.00$ & $18.0 \pm 1.8$ & $158.0 \pm 9.5$ & $605.9 \pm 36.4$ & $89.3 \pm 5.4$ \\
 17 & $-2.09 \pm 0.02$ & $2.79 \pm 0.24$ & $4.37 \pm 0.37$ & $0.06 \pm 0.01$ & $18.8 \pm 1.9$ & $116.3 \pm 7.0$ & $514.2 \pm 30.9$ & $85.9 \pm 5.2$ \\
 18 & $-1.87 \pm 0.00$ & $4.81 \pm 0.41$ & $5.21 \pm 0.44$ & $0.13 \pm 0.02$ & $17.6 \pm 1.8$ & $41.3 \pm 2.5$ & $222.2 \pm 13.3$ & $23.0 \pm 1.4$ \\
 19 & $-2.31 \pm 0.06$ & $2.37 \pm 0.20$ & $3.41 \pm 0.29$ & $0.12 \pm 0.01$ & $4.3 \pm 0.4$ & $22.1 \pm 1.3$ & $77.6 \pm 4.7$ & $15.3 \pm 0.9$ \\
 20 & $-2.09 \pm 0.02$ & $3.66 \pm 0.31$ & $3.44 \pm 0.29$ & $0.08 \pm 0.01$ & $37.0 \pm 3.7$ & $112.0 \pm 6.7$ & $390.7 \pm 23.4$ & $48.5 \pm 2.9$ \\
 21 & $-1.86 \pm 0.01$ & $3.10 \pm 0.26$ & $4.21 \pm 0.36$ & $0.04 \pm 0.00$ & $12.5 \pm 1.2$ & $111.1 \pm 6.7$ & $474.1 \pm 28.4$ & $69.2 \pm 4.2$ \\
 22 & $-2.23 \pm 0.01$ & $3.35 \pm 0.28$ & $4.61 \pm 0.39$ & $0.15 \pm 0.02$ & $10.5 \pm 1.0$ & $36.2 \pm 2.2$ & $172.1 \pm 10.3$ & $24.9 \pm 1.5$ \\
 23 & $-2.64 \pm 0.10$ & $1.79 \pm 0.15$ & $4.01 \pm 0.34$ & $0.09 \pm 0.01$ & $4.2 \pm 0.4$ & $31.0 \pm 1.9$ & $128.9 \pm 7.7$ & $61.6 \pm 3.7$ \\
 24 & $-2.50 \pm 0.06$ & $5.51 \pm 0.47$ & $4.82 \pm 0.41$ & $0.11 \pm 0.01$ & $6.8 \pm 0.7$ & $46.2 \pm 2.8$ & $229.3 \pm 13.8$ & $17.3 \pm 1.0$ \\
\hline
 \textbf{Median\tablenotemark{c}} & $-2.2 \pm 0.04$ & $3.4 \pm 0.18$ &  $ 4.4 \pm 0.12$ & $0.13 \pm 0.01$ & $13 \pm 4.4$ & $46 \pm 8.8$ & $229 \pm 42$ & $30 \pm 6.4$ \\
\hline
\hline
Entire galaxy\tablenotemark{f} & $-2.20 \pm 0.03$ & $3.81 \pm 0.32$ & $4.80 \pm 0.41$ & $0.14 \pm 0.02$ & $33.1 \pm 3.3$ & $103.8 \pm 6.2$ & $503.5 \pm 30.2$ & $55.7 \pm 3.3$ \\
\hline
\hline
\enddata

\tablenotetext{a}{The \lya\ escape fraction measured in (Section \ref{sec:Sec_2_6_emission line ratio_LyA_escape}), assuming  Case B recombination.}
\tablenotetext{b}{Rest-frame equivalent width.}
\tablenotetext{c}{The multiple images of the leaking region as identified in Figure \ref{fig0}. The associated R.A. and Decl. is listed in Table \ref{tab3}.}
\tablenotetext{d}{The reported uncertainties are the standard error of the mean of the measured parameters (i.e., divided by $\sqrt{N}$).}
\tablenotetext{e}{The non-leaking regions as identified in Figure \ref{fig0}. The associated R.A. and Decl. is listed in Table \ref{tab3}.}
\tablenotetext{f}{The entire galaxy (Arc 3) as identified in green in Figure \ref{fig0}.}


\label{tab2}
\end{deluxetable*}


\section{Discussion} 
\label{sec:Sec4_discussion} 

\subsection{Extreme Properties of the Compact ($< 100$ pc) LyC-leaking Region} 
\label{sec:Sec4_1_LyC_extm_properties}
Our key result is that the LyC leaking region within the Sunburst Arc has dramatically different properties than the rest of the galaxy.
The LyC leaking clump is an extremely compact star-forming region with very blue UV-continuum slope ($\beta \sim -2.9$), high ionization (i.e., \OIII /\OII\ $\sim 11$), high \Lya\ escape fraction (\Lyaesc\ $\sim 0.3$), and high oxygen EW(\OIII ) $\sim 1100 \ \angstrom$ (Figures \ref{fig1} and \ref{fig2} and Table \ref{tab2}).
The LyC leaking region is only a small unresolved (< 100 pc) piece of the entire galaxy (Figure~\ref{fig0}), and it exhibits physical conditions that are clearly most extreme over all other (non-leaking) star-forming regions.

Morphologically, such a compact shape of the leaking region seems related to the presence of dense star clusters and suggests the importance of concentrated star formation to the escape process of LyC radiation.
Indeed, (non-lensed) LyC leakers exhibit markedly similar compact star formation at all redshifts ($0.02 < z < 3.5$) \citep[e.g.,][]{berg06,bort14,vanz16,izot16a,izot18a}.
Due to the concentrated star formation, the leakers show high star formation surface density, $\Sigma{\rm{SFR}} \gtrsim 1 M_{\odot} \ {\rm yr^{ -1} \ kpc^{-2}}$ \citep{izot16a,izot18a,kim20,kim21,flur22} and a significant correlation between LyC escape fraction and UV-continuum size \citep{flur22b}.

Unsurprisingly, we also see markedly high ionization state of gas (i.e., based on high \OIII /\OII\ of $\simeq 11$ and \OIII /\Hb\ $\simeq 7$, Figure \ref{fig1}) of the leaking region.
A well-known tracer for ionization state of gas \citep[e.g.,][]{kewl02,naka14,kewl19,naka22}, the \OIII /\OII\ line ratio of $\simeq 11$ corresponds to very high ionization parameter of ${\rm{log}} U$ $\simeq -2$ based on the photoionization models of \cite{kewl19} and \cite{naka22} using the best-fit metallicity of 0.5$Z_{\odot}$ for the leaking region \citep{chis19}. The high log $U$ $\simeq -2$ strongly indicates the presence of highly ionized, low $N$(\HI ) along the line of sight toward the leaking region. 

Such a high ionization parameter of the leaking region is notably comparable to those of `super star clusters' found in the local universe \citep[e.g.,][]{inde09,jame16,mich17,mich19,leit18}.
Given the high ionization parameter and the \textit{intrinsically} small size ($r \lesssim 50$ pc) of the leaking region after correcting for lensing magnification \citep{vanz22,shar22}, it is very likely  that the leaking region harbours an actively forming super star cluster(s).

\begin{figure}
\includegraphics[width=\linewidth]{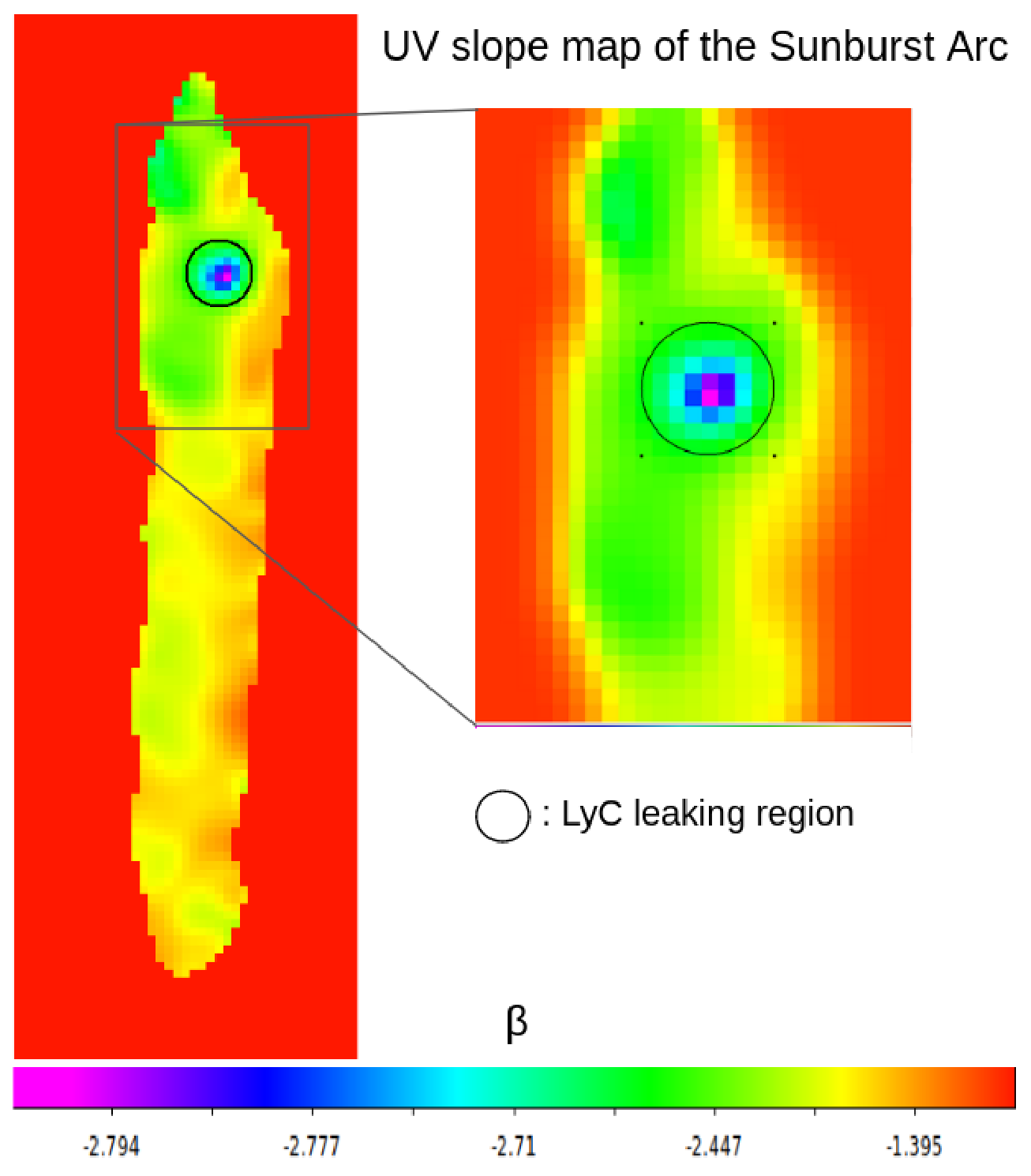}
\caption{UV slope map of the entire galaxy (i.e., `Arc 3' in Figure \ref{fig0}) and the zoom-in view of the leaking region shows the distinctly blue UV slope of the leaking region compared to any other non-leaking regions of the galaxy.
Consistent with previous figures, the $\beta$ map suggests that the escape of LyC photons in the Sunburst Arc is primarily driven by a local, compact starburst region that is likely harbouring a young star cluster(s) (Section \ref{sec:Sec4_1_LyC_extm_properties}).
\label{fig3_5_UV_slope_map}}
\end{figure}

These extreme properties of the leaking region are remarkably consistent with other, previously analyzed properties of the leaking region such as the strong highly ionized gas outflows \citep{main22} as well as the triple-peak profile of the \Lya\ emission line \citep{rive17}.
All of these properties of the leaking region show a coherent picture of a LyC escape mechanism where LyC photons escape from a particular extremely star-forming compact region (e.g., super star clusters) within a galaxy, rather than the entire galaxy.
Such a localized LyC escape process is also consistent with the trend shown in the UV slope map of the entire Sunburst galaxy (using the ``Arc 3'' image) in Figure \ref{fig3_5_UV_slope_map} where the leaking region distinctly features very blue $\beta$ unlike other non-leaking regions.

\subsection{Small Ionized Channels for the Escape of LyC and Nearly "Pure" Stellar Light}
\label{sec:Section4_2_UV_slope_Pure_stellar}
The leaking region's UV-continuum slope ($\beta = -2.9 \pm 0.1$, Figure \ref{fig1}) is extremely blue, compared to the typical slopes ($-2.5 \lesssim \beta \lesssim 0$) found in star-forming galaxies across redshift (0 $< z < 9$) \citep[e.g.,][]{meur99,hath08,dunl12,redd18,bhat21,chis22,topp22}.
How could such blue $\beta$ $\simeq -3$ be produced in the leaking region?
Theoretically, the very blue UV slope indicates the presence of extreme stellar populations characterized by little/no dust, young age, and low metallicity with a high ionizing photon escape fraction \citep{leit99,scha03,zack13,chis22,marq22}.
This would mean that ionizing photons are produced from young hot O (and late B type)-stars and the photons escape through empty space with little dust to avoid dust obscuration. Therefore, the leaking region's $\beta$ of $-2.9$ seems to strongly indicate that we are seeing the uninterrupted ``pure'' UV stellar light of these ionizing hot stars from the leaking region.

\begin{figure}
\includegraphics[width=\linewidth]{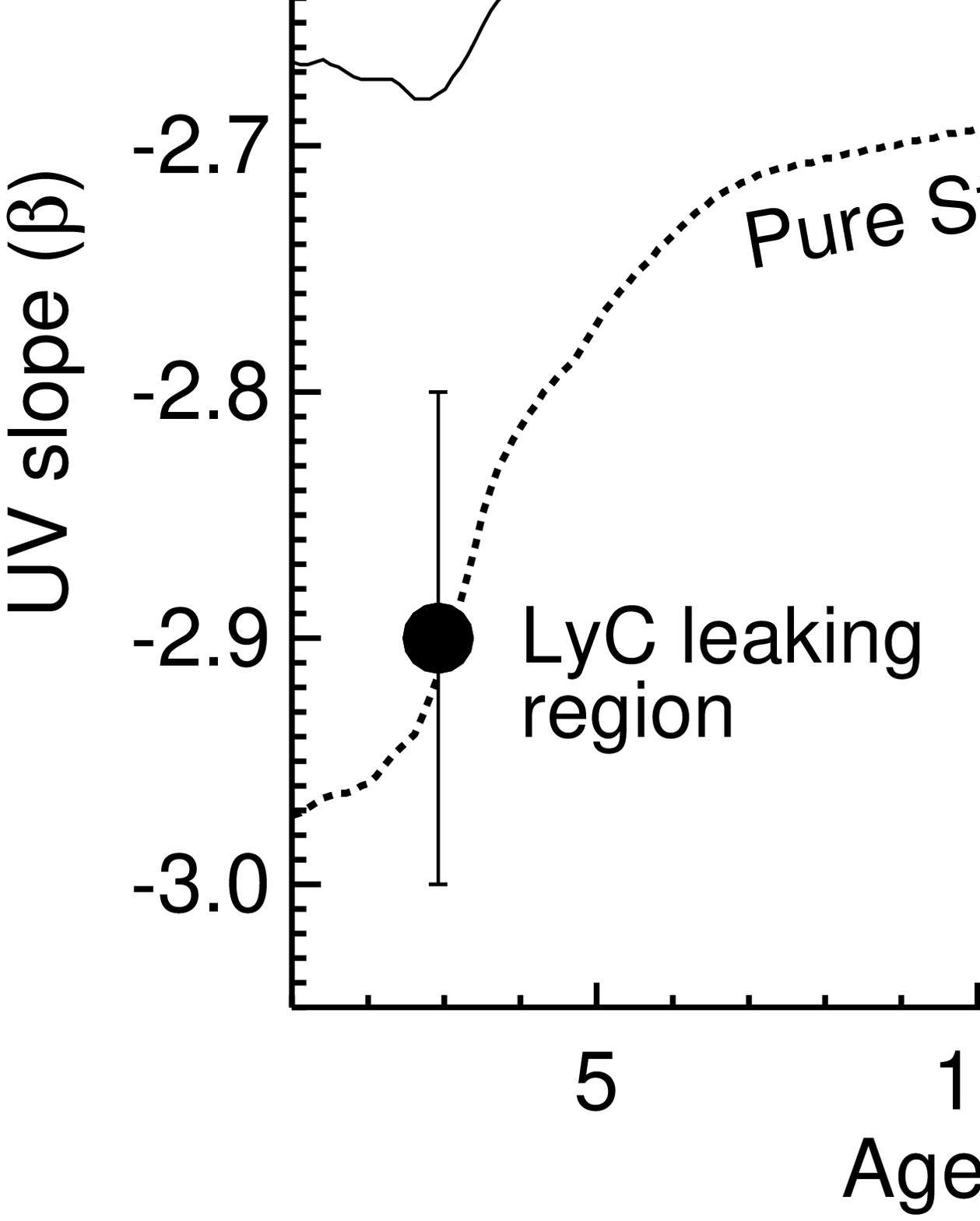}
\caption{The comparison of the leaking region's UV slope with starburst population models suggests that the leaking region's UV light is dominated by ``pure'' stellar light (i.e., minimal contribution from nebular continuum emission and extinction by dust).
The age ($\sim 3$ Myr) of the leaking region is adopted from the Far-UV SED modelling in \cite{chis19}.
Both `stellar+nebular' and `pure stellar' models assume the same constant star formation history and sub-solar metallicity of $0.4\ Z_{\odot}$ which is similar to the metallicity derived from \cite{chis19}.
The starburst population models are adopted from \cite{scha03}.
The leaking region's $\beta$ consistent with the ``pure stellar'' model suggests that the leaking region's UV emission nearly exclusively consists of uninterrupted direct star light, with minimal contributions from nebular continuum emission and negligible dust extinction. Such direct star light from the leaking region indicates the presence of small ionized channels through which LyC photons escape. (Section \ref{sec:Section4_2_UV_slope_Pure_stellar}). 
\label{fig4}}
\end{figure}

Indeed, the comparison of the leaking region's UV slope with the modelled slopes for low metallicity young stellar populations \citep{scha03} shows that their slopes are markedly consistent within uncertainties.
The leaking region's $\beta = -2.9 \pm 0.1$ closely matches with the modelled $\beta$s for sub-solar metallicity ($Z=0.4 Z_{\odot}$) young ($\lesssim 5$ Myr) stellar populations as shown in Figure \ref{fig4}. 
Furthermore, the comparison shows that only the most extreme models containing approximately ``pure'' stellar light---i.e., minimal contribution of nebular continuum emission---are still able to predict $\beta$ values that match with the observed $\beta$ of the leaking region. We checked that even for very low metallicity starburst population models ($0.02-0.2 Z_{\odot}$) only ``pure'' stellar light models are able to predict the blue $\beta$s comparable to that of the leaking region.
Such remarkable agreement in UV slopes between the leaking region and the ``pure'' stellar light models suggests that there is an exceptionally low \HI\ column density channel through which the unabsorbed ``direct'' stellar light including LyC photons are able to escape without interruptions by neutral hydrogen gas and dust contents.

These results point to a physical picture in which strong stellar feedback by a cluster of massive stars in the leaking region creates some highly ionized cavities of low neutral hydrogen and/or little dust. Through these ``pencil-beam'' cavities, ``direct'' stellar light, including LyC photons, are able to escape into the IGM.
The cavities would also help \Lya\ photon escape by reducing the scatterings with the surrounding gas, in agreement with the higher \Lyaesc\ (Figure \ref{fig2}) and the presence of central, narrow \Lya\ peak profiles \citep{rive17} in the leaking region than in the non-leaking regions.
In this viewing geometry, we may be staring ``down-the-barrel'' of the inner hot \HII \ region surrounding the ionizing star cluster; In that inner \HII \ region, highly ionized gas produces strong emission lines (such as \OIII ) while nebular continuum is relatively weak as gas is mostly ionized suggesting that the free-bound process is unlikely to contribute to the nebular continuum \citep{moll09}.
Without the extremely high magnification from strong lensing we would be not be able to separate the leaking region from the non-leaking regions and pin down this specific escape channel(s). We will further discuss the viewing geometry of LyC escape in Section \ref{sec:Section4_4_viewing_geometry}.

\subsection{LyC Escape Tracks with \Lya\ Escape On Sub-galactic Scales in the Sunburst Arc}
\label{sec:Section4_3_LyC_LyA}
The escape of \lya\ photons is one of the most compelling indirect indicators for LyC leakage \citep[e.g,.][]{behr14,verh15,rive17,izot21,kimm22}.
Both \Lya\ and LyC photons require low column density \HI\ gas to escape, although the precise escape fractions of \Lya\ and LyC photons are expected to substantially differ due to the larger interaction cross-section of \Lya\ with \HI\ atoms---a consequence of the resonant scattering nature of \Lya\ photons \citep[e.g.,][]{neuf91,gron16}.
Consistent with these expectations, we find that the leaking region shows higher \Lyaesc\ compared to the non-leaking regions; the mean \Lyaesc\ of the leaking and the non-leaking regions are 0.3 $\pm\ 0.03$ and 0.13 $\pm\ 0.07$, respectively (Figure \ref{fig2} and Table \ref{tab2}). Note that \Lyaesc\ here is measured on individual star-forming region scales and thus should be interpreted as a ``local'' line-of-sight \lya\ escape fraction as described in Section \ref{sec:Sec_2_6_emission line ratio_LyA_escape}.

It is especially interesting to see in the Sunburst Arc that both the \Lyaesc\ and LyC escape fraction vary dramatically on \textit{sub-galactic} scales.
This reinforces the fact that geometric factors such as low HI gas channels along the line of sight are critically important to the escape mechanisms for both.
Ongoing investigations on the \lya\ emission profiles of the leaking region and the non-leaking regions in the Sunburst Arc will further elucidate the interdependence between LyC and \lya\ escape mechanisms (Owens et al. and Rivera-Thorsen et al., in prep.).

\subsection{The Viewing Geometry for LyC Escape from Star Clusters}
\label{sec:Section4_4_viewing_geometry}
The Sunburst Arc is a clear example of a LyC escape process that is driven by a specific compact star-forming region within a galaxy.
The compact and vigorously star-forming region is harbouring a super star cluster considering its small, compact size with extreme properties (Figures \ref{fig1} and \ref{fig2} and Table \ref{tab2}).
The super star cluster 1) produces an enormous amount of ionizing photons and 2) likely plays a key role in enabling the photons to escape into the IGM through specific lines of sight where there is little nebular continuum emission and dust screening.
These unique insights into the locally regulated (i.e., not galaxy-wide) LyC escape processes by a prominent star cluster are only achievable by accurately pinning down the particular region of LyC leakage within the galaxy. 
Such observations require the unique combination of a strongly gravitationally lensed LyC leaker (such as the Sunburst Arc) and the sharp angular resolution of HST.

The multiple-lensed images of the leaking region (i.e., the numbered knots in Figure \ref{fig0}) also allow us to investigate whether the escape of LyC is directional or isotropic (i.e., the viewing geometry). Even among the separate lensed images of the LyC leaking region in the Sunburst Arc we see significant scatters among both the physical properties (i.e., Figures \ref{fig1} and \ref{fig2}) and the LyC escape fraction \citep{rive19}. 
From the lens model of the system we know that the gravitational deflection that produces each of these lensed images results in a slightly different viewing angle and a different magnification factor for each image \citep[see Figures 7 and 14 in ][]{shar22}. Thus, the variations in the physical properties measured from the lensed images indicate that the geometry of the leaking region is very likely anisotropic such that its measured properties depend on which line of sight we look through (e.g., patchy \HI\ clouds and cavities in the ISM).

There are substantial variations among the properties measured from the leaking region's lensed images (i.e., the scattered  distribution of the black data points in Figures 2 and 3).
The variations in $\beta$ ($-3 \lesssim \beta \lesssim -2.7$), \OIII /\OII\ ($7 \lesssim \OIII /\OII\ \lesssim 15$), and \Lyaesc\ ($0.25 \lesssim$ \Lyaesc\ $\lesssim 0.35$) are larger than measurement uncertainties and are likely related to the details of lensing configuration, including slightly different viewing geometries and different lensing magnifications of the leaking region images. Qualitatively, we see trends between the variations in the physical properties and lensing magnification such that highly magnified images (e.g., clumps 10 and 2) tend to show more extreme properties such as bluer UV slope and higher \OIII /\OII\ ionization state.

This is consistent with the leaking LyC radiation emerging from a narrow, ``pencil-beam'' channel that is, at best, barely resolved. In this physical picture, some lensed images provide a cleaner view of the escape channel, while at the same time the more highly magnified images provide a cleaner, better-isolated view of the LyC leaking line of sight than the lower magnification images which are likely more blended with surrounding non-leaking regions. A quantitative assessment of these lensing effects requires a careful forward-modelling analysis, which is beyond the scope of this paper.

\begin{figure*}
\includegraphics[width=\linewidth]{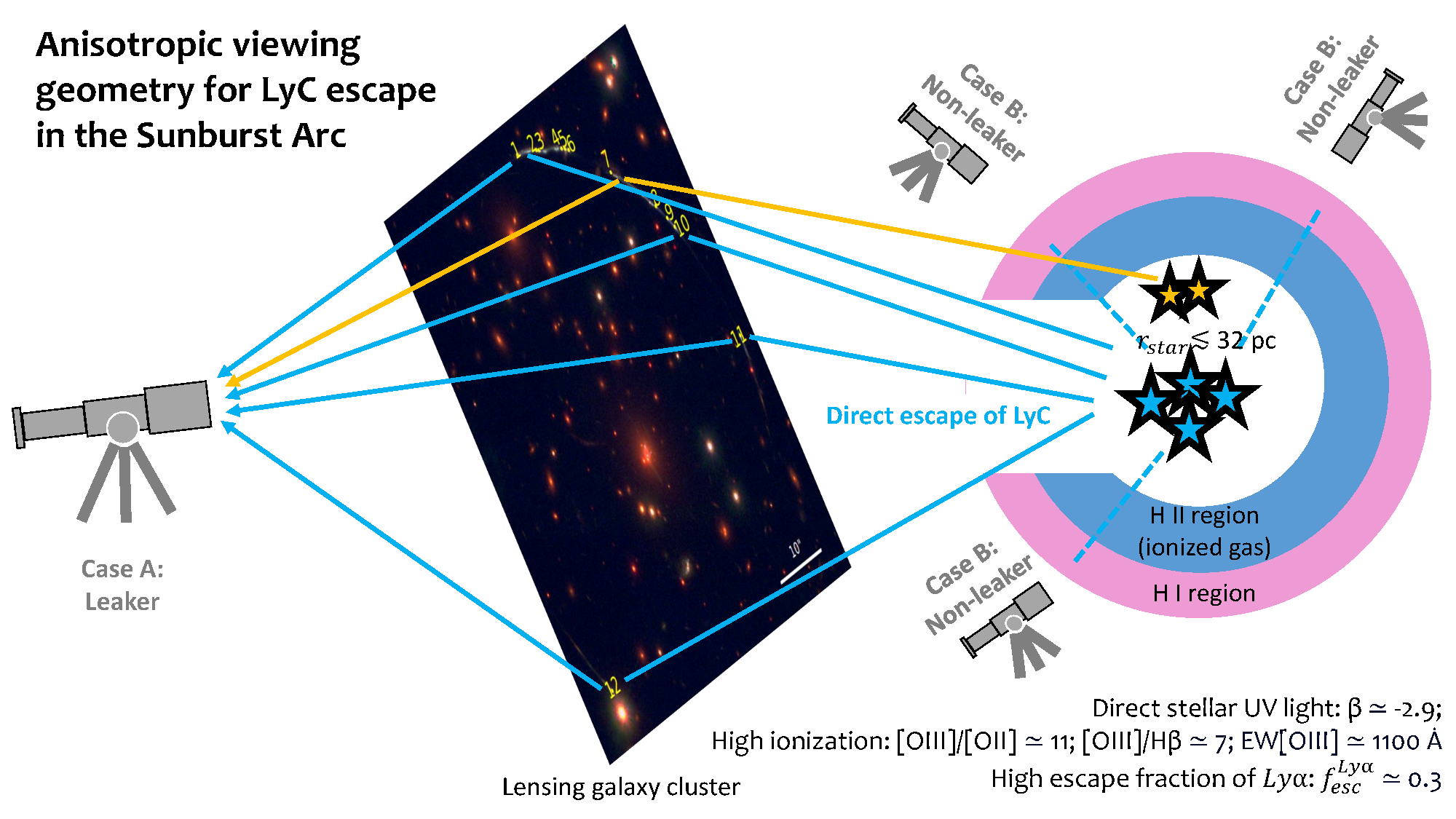}
\caption{The illustrative diagram on the anisotropic viewing geometry for LyC escape in the Sunburst Arc. Case A is the ``pencil beam'' viewing geometry suggested for the galaxy, where LyC photons directly escape from the ionizing star cluster through a highly ionized, low $N$(\HI ) hole(s) that opens towards our line of sight.
In this viewing geometry, we may be staring ``down-the-barrel'' of the inner hot \HII \ region of the ionizing star cluster, enabling us to observe LyC photons and pristine stellar light that are minimally attenuated by neutral hydrogen and dust, as indicated by the very blue $\beta \simeq -2.9$ of the leaking region (Section \ref{sec:Section4_4_viewing_geometry}).
This anisotropic LyC escape process through specific opening angles implies that if we had viewed the galaxy from most other angles, we may not see escaping ionizing photons and thus, identify it as a non-leaker, which is depicted as Case B.
Thanks to strong lensing magnification, our analysis isolates the physical properties of the leaking region (blue stars) from the non-leaking regions (orange stars).
The size ($r_{\rm star} \lesssim 32$ pc) of the ionizing star cluster is an upper limit due to the HST's spatial resolution and is estimated from the strong lens model analysis of \cite{shar22}.
\label{fig7}}
\end{figure*}

The systematic variations found in this study clearly suggest that the structure of LyC leaking region and the associated LyC escape process are highly directional (aniostropic) depending on a specific line of sight.
This means that we are viewing the galaxy from a very privileged vantage point through pencil-beam channels.
If we had viewed it from most other angles, we may not see ionizing escape. 
Such anisotropic viewing geometry for LyC escape in the Sunburst Arc is illustrated in Figure \ref{fig7}.
While the LyC escape fraction along our line of sight is very high $\sim 40 \%$, the galaxy-integrated global escape fraction is likely lower considering the small projected area of the leaking region within the galaxy.
Combined with such privileged sight-lines, the global escape fraction is on the order of $5 \%$ given that the leaking region accounts for $\sim 17 \%$ of the galaxy's non-ionizing UV continuum light (based on the F555W flux).
If the ``pencil-beam'' viewing geometry seen in the Sunburst Arc is a common way of escaping in star-forming galaxies, it is likely that field searches are missing substantial populations of LyC leakers for which leaking channels are misaligned with our viewing angle.

\subsection{Implications for Reionization}
\label{sec:Section4_5_Implications_for_Reionization}
One clear implication of our analysis is that the LyC escape process is a highly local process, and individual young, dense star clusters play a key role in facilitating the escape of LyC photons. The unique magnified view of LyC leakage provided in the Sunburst Arc reveals that one such LyC leaking star cluster is characterized by an extremely blue UV slope and high ionization state, while the non-leaking regions of the galaxy have properties consistent with a typical star-forming galaxy. In fact, the integrated, galaxy-averaged properties can be quite unremarkable, as demonstrated by the integrated properties measured for the Sunburst Arc using the low-magnification complete image of the galaxy (``Arc 3'' in Figure~\ref{fig0} and the blue diamonds in Figures \ref{fig1} and \ref{fig2}).

A second implication is that LyC escape is, in at least some cases, a highly anisotropic process (i.e., Figure \ref{fig7}). The extreme physical conditions that are associated with LyC escape only occur across a very small fraction of the surface of the Sunburst Arc, implying that the ionizing radiation is escaping through channels that subtend a very small solid angle (i.e., through channels that are long and thin). If pencil-beam channels are a common mode of LyC escape then it follows that detecting LyC escape from any single galaxy will be likely strongly viewing angle dependent, such that measurements of LyC escape in individual galaxies cannot safely be used to measure the volume-averaged escaping ionizing radiation due to viewing angle effects.

We can draw two important conclusions about LyC leaking galaxies: 1) their galaxy-integrated properties need not be extreme and 2) their escaping LyC radiation is likely to be highly anisotropic and viewing-geometry dependent (i.e., favorable sightlines). The combination of locally regulated LyC escape mechanisms by super star clusters and pencil-beam geometries for channels of escaping ionizing radiation implies that it should be quite difficult to draw robust conclusions about galaxies' escaping LyC radiation using indirect indicators based on \textit{integrated, galaxy-averaged properties}. Notably, these pencil-beam channels driven by small, individual star clusters provide a natural explanation for why studies analyzing the integrated galaxy properties associated with LyC escape have found significant scatters (or weak trends) between those integrated galaxy properties and LyC escape fraction \citep[e.g.,][]{chis18,izot21,flur22,saxe22,sald22,seiv22}.






\section{Summary and Conclusions} 
\label{sec:Sec_5_Summary_conclusions}

We investigate the physical conditions for LyC escape by isolating and measuring the key properties of a leaking region in a strongly lensed, LyC emitter at $z=2.37$ (aka, Sunburst Arc).
Thanks to high magnification from strong lensing, this galaxy reveals the exceptionally small scale (tens of parsecs) physics of LyC escape; its lensing-magnified images reveal that only one compact star-forming region emits ionizing photons while the other regions do not (Figure \ref{fig0}).
Analyzing the HST's sharp images, we spatially resolve the properties of the leaking region ($<$ 100 pc) and compare with the non-leaking regions as well as the entire galaxy-averaged properties. 

Our primary conclusions are summarized below:
\begin{itemize}
 
    \item The galaxy reveals significant variations among the physical properties on the spatial scales of individual star-clusters. Notably, the compact, small ($<$ 100 pc) LyC leaking region likely harbouring a young star cluster exhibits the most extreme physical properties: a very blue UV-continuum slope ($\beta = -2.9 \pm 0.1$), high ionization state (\OIII /\OII $= 11 \pm 3$ and \OIII /\Hb\ $=6.8 \pm 0.4$), strong oxygen emission (EW\OIII\ $= 1095 \pm 40 \ \angstrom$), and high Lyman-$\alpha$ escape fraction (\Lyaesc = $0.3 \pm 0.03$) (Figures \ref{fig1} and \ref{fig2}).
    Such extreme properties are not found in any non-leaking regions of the galaxy (Section \ref{sec:Sec4_1_LyC_extm_properties}).
    
    \item The leaking region's blue $\beta \simeq -2.9$ comparison with the starburst population models indicates that its UV emission consists of nearly ``pure'' stellar light with minimal contamination from surrounding nebular continuum emission and dust extinction (Figure \ref{fig4}). This suggests a direct escape of LyC photons from the ionizing star cluster through an ionized, low $N$(\HI ) opening channel(s) (Section \ref{sec:Section4_2_UV_slope_Pure_stellar}).

    \item The leaking region shows higher \Lyaesc\ and higher EW(\lya ) compared to the non-leaking regions (that is, \Lyaesc\ $\approx$ 0.3 vs. 0.13; EW(\Lya ) $\approx$ 43 $\angstrom$ vs. 13 $\angstrom$ as in Figure \ref{fig2} and Table \ref{tab2}). This suggests a similar escape process of LyC and \Lya\ photons such as preferred low $N$(\HI ) channels on \textit{sub-galactic} scales, although the detailed escape pathways can differ from each other due to the larger interaction cross-section of \lya\ with neutral hydrogen  \citep[e.g.,][]{neuf91,verh15} (Section \ref{sec:Section4_3_LyC_LyA}).

    \item Combined with the different lensing magnification factor and a slightly different viewing angle among the lensed images of the leaking region, the variations found among the physical properties of the lensed leaking region (i.e., the scattered distribution of the black data points in Figures \ref{fig1} and \ref{fig2}) clearly suggest that the structure of LyC leaking region and the associated LyC escape process are highly aniostropic depending on specific line of sight (Section \ref{sec:Section4_4_viewing_geometry}).
    This means that we are viewing the galaxy from a very privileged vantage point through ``pencil-beam'' opening angle. If we had viewed it from most other angles, we may not see ionizing escape (Figure \ref{fig7}). Such anisotropic LyC escape from a small, dense, star cluster in the Sunburst Arc is only identifiable due to the high magnification from strong lensing, which provides a zoomed-in view of the narrow LyC escape channel. With presently available observational facilities, spatially resolved studies of LyC leakage are only possible with distant, strongly lensed systems.
    
    \item The integrated \textit{galaxy} properties are not as extreme as those of the leaking region (i.e., the blue diamonds in Figures \ref{fig1} and \ref{fig2}), which show $\beta \simeq -2.2$, \OIII /\OII $\ \simeq 4$, and \OIII /\Hb\ $\simeq 5$), EW\OIII\ $\simeq 500 \ \angstrom$, and \Lyaesc\ $\simeq 0.13$. The galaxy properties are rather consistent with those typical of Lyman Break galaxies at similar redshifts \citep[e.g.,][]{hath08} and of local Green Pea galaxies \citep[e.g.,][]{henr15,yang17,izot16,izot21,flur22,chis22}. 
    However, the Sunburst Arc is a clear example of a localized, anisotropic process of LyC escape driven by a compact starburst region harbouring young star cluster(s), rather than the entire galaxy contributing to LyC escape.
    This implies that the true, volume-averaged escape fraction of ionizing radiation estimated for individual galaxies may often be subject to large systematic uncertainties due to random line-of-sight effects. It is possible---even likely---that unlensed star-forming galaxies like the Sunburst Arc would have integrated properties that are not indicative of LyC escape, but could in fact, be releasing significant ionizing radiation into the IGM (Section \ref{sec:Section4_5_Implications_for_Reionization}). These line-of-sight variations are also a natural explanation for the significant scatters between galaxy properties and LyC escape fraction in the LyC leakers \citep[e.g.,][]{izot22,saxe22,flur22b}.
     
\end{itemize}

To summarize, our results isolate the physical conditions for LyC escape on very small, sub-galactic scales and suggest an anisotropic LyC escape process driven by a compact young star cluster in galaxies.
If the Sunburst Arc is representative of how LyC escapes typical star-forming galaxies, then strong lensing is an essential tool for revealing how these galaxies contribute LyC photons to reionization by isolating the physical conditions of a specific LyC leaking region with boosted spatial magnification. Importantly, our results call for a more sophisticated reionization modelling that accounts for such directional (anisotropic) LyC escape mode of galaxies.

Going forward, the upcoming JWST IFU study on the Sunburst Arc (GO: 2555; PI: T Rivera-Thorsen) will further characterize the ISM and stellar properties of the leaking region by measuring key quantities such as the gas kinematics and dust geometry on spatially-resolved scales.

\vspace{5mm}
We thank the referee for constructive comments that improved the quality of the manuscript. The data presented in this paper were obtained from the Mikulski Archive for Space Telescopes (MAST) at the Space Telescope Science Institute. 
The specific observations analyzed can be accessed in MAST: \dataset[10.17909/5bmh-yx42]{https://doi.org/10.17909/5bmh-yx42}. 
Support for HST-GO-15101, HST-GO-15418, HST-GO-15377, and HST-GO-15949 was provided by NASA through grants from the Space Telescope Science Institute, which is operated by the Associations of Universities for Research in Astronomy (AURA), Incorporated, under NASA contract NAS5-26555.

\software{Astrodrizzle (Hack et al. 2012), STSYNPHOT/SYNPHOT (Lim 2019)}


\begin{thebibliography}{}

\bibitem[\protect\citeauthoryear{Abadi et al.}{1999}]{abad99} Abadi, M.~G., Moore, B., \& Bower, R.~G.\ 1999, \mnras, 308, 947

\bibitem[\protect\citeauthoryear{Behrens et al.}{2014}]{behr14} Behrens, C., Dijkstra, M., \& Niemeyer, J.~C.\ 2014, \aap, 563, A77

\bibitem[\protect\citeauthoryear{Bergvall \& Olofsson}{1986}]{berg86} Bergvall, N. \& Olofsson, K.\ 1986, \aaps, 64, 469

\bibitem[\protect\citeauthoryear{Bergvall et al.}{2006}]{berg06} Bergvall, N., Zackrisson, E., Andersson, B.-G., et al.\ 2006, \aap, 448, 513

\bibitem[\protect\citeauthoryear{Bhatawdekar \& Conselice}{2021}]{bhat21} Bhatawdekar, R. \& Conselice, C.~J.\ 2021, \apj, 909, 144

\bibitem[\protect\citeauthoryear{Borthakur et al.}{2014}]{bort14} Borthakur, S., Heckman, T.~M., Leitherer, C., et al.\ 2014, Science, 346, 216

\bibitem[\protect\citeauthoryear{Bouwens et al.}{2010}]{bouw10} Bouwens, R.~J., Illingworth, G.~D., Oesch, P.~A., et al.\ 2010, \apjl, 708, L69

\bibitem[\protect\citeauthoryear{Calzetti et al.}{2000}]{calz00} Calzetti, D., Armus, L., Bohlin, R.~C., et al.\ 2000, \apj, 533, 682

\bibitem[\protect\citeauthoryear{Calzetti et al.}{1994}]{calz94} Calzetti, D., Kinney, A.~L., \& Storchi-Bergmann, T.\ 1994, \apj, 429, 582

\bibitem[\protect\citeauthoryear{Cardelli et al.}{1989}]{card89} Cardelli, J.~A., Clayton, G.~C., \& Mathis, J.~S.\ 1989, \apj, 345, 245

\bibitem[\protect\citeauthoryear{Chisholm et al.}{2018}]{chis18} Chisholm, J., Gazagnes, S., Schaerer, D., et al.\ 2018, \aap, 616, A30. doi:10.1051/0004-6361/201832758

\bibitem[\protect\citeauthoryear{Chisholm et al.}{2019}]{chis19} Chisholm, J., Rigby, J.~R., Bayliss, M., et al.\ 2019, \apj, 882, 182

\bibitem[\protect\citeauthoryear{Chisholm et al.}{2022}]{chis22} Chisholm, J., Saldana-Lopez, A., Flury, S., et al.\ 2022, arXiv:2207.05771

\bibitem[\protect\citeauthoryear{Dahle et al.}{2016}]{dahl16} Dahle, H., Aghanim, N., Guennou, L., et al.\ 2016, \aap, 590, L4

\bibitem[\protect\citeauthoryear{Diego et al.}{2022}]{dieg22} Diego, J.~M., Pascale, M., Kavanagh, B.~J., et al.\ 2022, \aap, 665, A134

\bibitem[\protect\citeauthoryear{Dijkstra}{2019}]{dijk19} Dijkstra, M.\ 2019, Saas-Fee Advanced Course, 46, 1. doi:10.1007/978-3-662-59623-4\_1

\bibitem[\protect\citeauthoryear{Dopita \& Sutherland}{2003}]{dopi03} Dopita, M.~A. \& Sutherland, R.~S.\ 2003, Astrophysics of the diffuse universe, Berlin, New York: Springer, 2003. Astronomy and astrophysics library, ISBN 3540433627

\bibitem[\protect\citeauthoryear{Dunlop et al.}{2012}]{dunl12} Dunlop, J.~S., McLure, R.~J., Robertson, B.~E., et al.\ 2012, \mnras, 420, 901

\bibitem[\protect\citeauthoryear{Fan et al.}{2006}]{fan06} Fan, X., Carilli, C.~L., \& Keating, B.\ 2006, \araa, 44, 415

\bibitem[\protect\citeauthoryear{Finkelstein et al.}{2019}]{fink19} Finkelstein, S.~L., D'Aloisio, A., Paardekooper, J.-P., et al.\ 2019, \apj, 879, 36

\bibitem[\protect\citeauthoryear{Florian et al.}{2021}]{flor21} Florian, M.~K., Rigby, J.~R., Acharyya, A., et al.\ 2021, \apj, 916, 50

\bibitem[\protect\citeauthoryear{Flury et al.}{2022a}]{flur22} Flury, S.~R., Jaskot, A.~E., Ferguson, H.~C., et al.\ 2022a, \apj, 930, 126

\bibitem[\protect\citeauthoryear{Flury et al.}{2022b}]{flur22b} Flury, S.~R., Jaskot, A.~E., Ferguson, H.~C., et al.\ 2022b, \apjs, 260, 1

\bibitem[\protect\citeauthoryear{Gazagnes et al.}{2020}]{gaza20} Gazagnes, S., Chisholm, J., Schaerer, D., et al.\ 2020, \aap, 639, A85

\bibitem[\protect\citeauthoryear{Gronke et al.}{2016}]{gron16} Gronke, M., Dijkstra, M., McCourt, M., et al.\ 2016, \apjl, 833, L26

\bibitem[\protect\citeauthoryear{Hathi et al.}{2008}]{hath08} Hathi, N.~P., Malhotra, S., \& Rhoads, J.~E.\ 2008, \apj, 673, 686

\bibitem[\protect\citeauthoryear{Henry et al.}{2015}]{henr15} Henry, A., Scarlata, C., Martin, C.~L., et al.\ 2015, \apj, 809, 19

\bibitem[\protect\citeauthoryear{Indebetouw et al.}{2009}]{inde09} Indebetouw, R., de Messi{\`e}res, G.~E., Madden, S., et al.\ 2009, \apj, 694, 84

\bibitem[\protect\citeauthoryear{Izotov et al.}{2022}]{izot22} Izotov, Y.~I., Chisholm, J., Worseck, G., et al.\ 2022, \mnras, 515, 2864

\bibitem[\protect\citeauthoryear{Izotov et al.}{2016a}]{izot16a} Izotov, Y.~I., Orlitov{\'a}, I., Schaerer, D., et al.\ 2016, \nat, 529, 178. doi:10.1038/nature16456

\bibitem[\protect\citeauthoryear{Izotov et al.}{2016b}]{izot16} Izotov, Y.~I., Schaerer, D., Thuan, T.~X., et al.\ 2016, \mnras, 461, 3683

\bibitem[\protect\citeauthoryear{Izotov et al.}{2018a}]{izot18a} Izotov, Y.~I., Schaerer, D., Worseck, G., et al.\ 2018, \mnras, 474, 4514. doi:10.1093/mnras/stx3115

\bibitem[\protect\citeauthoryear{Izotov et al.}{2018b}]{izot18} Izotov, Y.~I., Worseck, G., Schaerer, D., et al.\ 2018, \mnras, 478, 4851

\bibitem[\protect\citeauthoryear{Izotov et al.}{2021}]{izot21} Izotov, Y.~I., Worseck, G., Schaerer, D., et al.\ 2021, \mnras, 503, 1734

\bibitem[\protect\citeauthoryear{James et al.}{2016}]{jame16} James, B.~L., Auger, M., Aloisi, A., et al.\ 2016, \apj, 816, 40

\bibitem[\protect\citeauthoryear{Ji et al.}{2020}]{ji20} Ji, Z., Giavalisco, M., Vanzella, E., et al.\ 2020, \apj, 888, 109

\bibitem[\protect\citeauthoryear{Keenan et al.}{2017}]{keen17} Keenan, R.~P., Oey, M.~S., Jaskot, A.~E., et al.\ 2017, \apj, 848, 12

\bibitem[\protect\citeauthoryear{Kewley \& Dopita}{2002}]{kewl02} Kewley, L.~J. \& Dopita, M.~A.\ 2002, \apjs, 142, 35

\bibitem[\protect\citeauthoryear{Kewley et al.}{2019}]{kewl19} Kewley, L.~J., Nicholls, D.~C., \& Sutherland, R.~S.\ 2019, \araa, 57, 511

\bibitem[\protect\citeauthoryear{Kim et al.}{2020}]{kim20} Kim, K., Malhotra, S., Rhoads, J.~E., et al.\ 2020, \apj, 893, 134

\bibitem[\protect\citeauthoryear{Kim et al.}{2021}]{kim21} Kim, K.~J., Malhotra, S., Rhoads, J.~E., et al.\ 2021, \apj, 914, 2

\bibitem[\protect\citeauthoryear{Kimm et al.}{2022}]{kimm22} Kimm, T., Bieri, R., Geen, S., et al.\ 2022, \apjs, 259, 21

\bibitem[\protect\citeauthoryear{Leitet et al.}{2013}]{leit13} Leitet, E., Bergvall, N., Hayes, M., et al.\ 2013, \aap, 553, A106

\bibitem[\protect\citeauthoryear{Leitet et al.}{2011}]{leit11} Leitet, E., Bergvall, N., Piskunov, N., et al.\ 2011, \aap, 532, A107

\bibitem[\protect\citeauthoryear{Leitherer et al.}{2016}]{leit16} Leitherer, C., Hernandez, S., Lee, J.~C., et al.\ 2016, \apj, 823, 64

\bibitem[\protect\citeauthoryear{Leitherer et al.}{2018}]{leit18} Leitherer, C., Byler, N., Lee, J.~C., et al.\ 2018, \apj, 865, 55

\bibitem[\protect\citeauthoryear{Leitherer et al.}{1999}]{leit99} Leitherer, C., Schaerer, D., Goldader, J.~D., et al.\ 1999, \apjs, 123, 3

\bibitem[\protect\citeauthoryear{Lim}{2019}]{lim19} Lim, P. L.\ 2019, Zenodo

\bibitem[\protect\citeauthoryear{Lopez et al.}{2020}]{lope20} Lopez, S., Tejos, N., Barrientos, L.~F., et al.\ 2020, \mnras, 491, 4442

\bibitem[\protect\citeauthoryear{Mainali et al.}{2022}]{main22} Mainali, R., Rigby, J.~R., Chisholm, J., et al.\ 2022, arXiv:2210.11575

\bibitem[\protect\citeauthoryear{Malkan \& Malkan}{2021}]{malk21} Malkan, M.~A. \& Malkan, B.~K.\ 2021, \apj, 909, 92

\bibitem[\protect\citeauthoryear{Marques-Chaves et al.}{2022}]{marq22} Marques-Chaves, R., Schaerer, D., {\'A}lvarez-M{\'a}rquez, J., et al.\ 2022, \mnras, 517, 2972

\bibitem[\protect\citeauthoryear{Me{\v{s}}tri{\'c} et al.}{2023}]{mest23} Me{\v{s}}tri{\'c}, U., Vanzella, E., Upadhyaya, A., et al.\ 2023, \aap, 673, A50

\bibitem[\protect\citeauthoryear{Meurer et al.}{1999}]{meur99} Meurer, G.~R., Heckman, T.~M., \& Calzetti, D.\ 1999, \apj, 521, 64

\bibitem[\protect\citeauthoryear{Micheva et al.}{2019}]{mich19} Micheva, G., Christian Herenz, E., Roth, M.~M., et al.\ 2019, \aap, 623, A145

\bibitem[\protect\citeauthoryear{Micheva et al.}{2017}]{mich17} Micheva, G., Oey, M.~S., Jaskot, A.~E., et al.\ 2017, \apj, 845, 165

\bibitem[\protect\citeauthoryear{Moll{\'a} et al.}{2009}]{moll09} Moll{\'a}, M., Garc{\'\i}a-Vargas, M.~L., \& Bressan, A.\ 2009, \mnras, 398, 451

\bibitem[\protect\citeauthoryear{Mostardi et al.}{2015}]{most15} Mostardi, R.~E., Shapley, A.~E., Steidel, C.~C., et al.\ 2015, \apj, 810, 107

\bibitem[\protect\citeauthoryear{Naidu et al.}{2020}]{naid20} Naidu, R.~P., Tacchella, S., Mason, C.~A., et al.\ 2020, \apj, 892, 109

\bibitem[\protect\citeauthoryear{Nakajima \& Ouchi}{2014}]{naka14} Nakajima, K. \& Ouchi, M.\ 2014, \mnras, 442, 900

\bibitem[\protect\citeauthoryear{Nakajima et al.}{2022}]{naka22} Nakajima, K., Ouchi, M., Xu, Y., et al.\ 2022, \apjs, 262, 3

\bibitem[\protect\citeauthoryear{Neufeld}{1991}]{neuf91} Neufeld, D.~A.\ 1991, \apjl, 370, L85

\bibitem[\protect\citeauthoryear{{\"O}stlin et al.}{2021}]{ostl21} {\"O}stlin G., Rivera-Thorsen T.~E., Menacho V., Hayes M., Runnholm A., Micheva G., Oey M.~S., et al., 2021, ApJ, 912, 155

\bibitem[\protect\citeauthoryear{Pascale et al.}{2023}]{pasc23} Pascale, M., Dai, L., McKee, C.~F., et al.\ 2023, arXiv:2301.10790

\bibitem[\protect\citeauthoryear{Ramambason et al.}{2020}]{ramam20} Ramambason, L., Schaerer, D., Stasi{\'n}ska, G., et al.\ 2020, \aap, 644, A21

\bibitem[\protect\citeauthoryear{Reddy et al.}{2018}]{redd18} Reddy, N.~A., Oesch, P.~A., Bouwens, R.~J., et al.\ 2018, \apj, 853, 56

\bibitem[\protect\citeauthoryear{Reddy et al.}{2016}]{redd16} Reddy, N.~A., Steidel, C.~C., Pettini, M., et al.\ 2016, \apj, 828, 107

\bibitem[\protect\citeauthoryear{Rivera-Thorsen et al.}{2017}]{rive17} Rivera-Thorsen, T.~E., Dahle, H., Gronke, M., et al.\ 2017, \aap, 608, L4

\bibitem[\protect\citeauthoryear{Rivera-Thorsen et al.}{2019}]{rive19} Rivera-Thorsen, T.~E., Dahle, H., Chisholm, J., et al.\ 2019, Science, 366, 738

\bibitem[\protect\citeauthoryear{Rivera-Thorsen et al.}{2015}]{rive15} Rivera-Thorsen, T.~E., Hayes, M., {\"O}stlin, G., et al.\ 2015, \apj, 805, 14

\bibitem[\protect\citeauthoryear{Robertson et al.}{2015}]{robe15} Robertson, B.~E., Ellis, R.~S., Furlanetto, S.~R., et al.\ 2015, \apjl, 802, L19

\bibitem[\protect\citeauthoryear{Rutkowski et al.}{2016}]{rutk16} Rutkowski, M.~J., Scarlata, C., Haardt, F., et al.\ 2016, \apj, 819, 81. doi:10.3847/0004-637X/819/1/81

\bibitem[\protect\citeauthoryear{Rutkowski et al.}{2017}]{rutk17} Rutkowski, M.~J., Scarlata, C., Henry, A., et al.\ 2017, \apjl, 841, L27

\bibitem[\protect\citeauthoryear{Saldana-Lopez et al.}{2022}]{sald22} Saldana-Lopez, A., Schaerer, D., Chisholm, J., et al.\ 2022, \aap, 663, A59

\bibitem[\protect\citeauthoryear{Saxena et al.}{2022}]{saxe22} Saxena, A., Pentericci, L., Ellis, R.~S., et al.\ 2022, \mnras, 511, 120

\bibitem[\protect\citeauthoryear{Schaerer}{2003}]{scha03} Schaerer, D.\ 2003, \aap, 397, 527

\bibitem[\protect\citeauthoryear{Seive et al.}{2022}]{seiv22} Seive, T., Chisholm, J., Leclercq, F., et al.\ 2022, \mnras, 515, 5556

\bibitem[\protect\citeauthoryear{Shapley et al.}{2016}]{shap16} Shapley, A.~E., Steidel, C.~C., Strom, A.~L., et al.\ 2016, \apjl, 826, L24

\bibitem[\protect\citeauthoryear{Sharon et al.}{2022}]{shar22} Sharon, K., Mahler, G., Rivera-Thorsen, T.~E., et al.\ 2022, arXiv:2209.03417

\bibitem[\protect\citeauthoryear{Simcoe et al.}{2013}]{simc13} Simcoe, R.~A., Burgasser, A.~J., Schechter, P.~L., et al.\ 2013, \pasp, 125, 270

\bibitem[\protect\citeauthoryear{Storey \& Zeippen}{2000}]{stor00} Storey, P.~J. \& Zeippen, C.~J.\ 2000, \mnras, 312, 813

\bibitem[\protect\citeauthoryear{Topping et al.}{2022}]{topp22} Topping, M.~W., Stark, D.~P., Endsley, R., et al.\ 2022, arXiv:2208.01610

\bibitem[\protect\citeauthoryear{Vanzella et al.}{2022}]{vanz22} Vanzella, E., Castellano, M., Bergamini, P., et al.\ 2022, \aap, 659, A2

\bibitem[\protect\citeauthoryear{Vanzella et al.}{2016}]{vanz16} Vanzella, E., de Barros, S., Vasei, K., et al.\ 2016, \apj, 825, 41

\bibitem[\protect\citeauthoryear{Vanzella et al.}{2018}]{vanz18} Vanzella, E., Nonino, M., Cupani, G., et al.\ 2018, \mnras, 476, L15

\bibitem[\protect\citeauthoryear{Verhamme et al.}{2015}]{verh15} Verhamme, A., Orlitov{\'a}, I., Schaerer, D., et al.\ 2015, \aap, 578, A7

\bibitem[\protect\citeauthoryear{Wang et al.}{2019}]{wang19} Wang, B., Heckman, T.~M., Leitherer, C., et al.\ 2019, \apj, 885, 57

\bibitem[\protect\citeauthoryear{Wofford et al.}{2013}]{woff13} Wofford, A., Leitherer, C., \& Salzer, J.\ 2013, \apj, 765, 118

\bibitem[\protect\citeauthoryear{Yang et al.}{2017}]{yang17} Yang, H., Malhotra, S., Gronke, M., et al.\ 2017, \apj, 844, 171

\bibitem[\protect\citeauthoryear{Yung et al.}{2020}]{yung20} Yung, L.~Y.~A., Somerville, R.~S., Finkelstein, S.~L., et al.\ 2020, \mnras, 496, 4574

\bibitem[\protect\citeauthoryear{Zackrisson et al.}{2013}]{zack13} Zackrisson, E., Inoue, A.~K., \& Jensen, H.\ 2013, \apj, 777, 39

\end{thebibliography}
\citestyle{aa}
\bibliographystyle{aasjournal}



\appendix
\section{R.A. and Decl. of the Regions analyzed}
Table \ref{tab3} lists the R.A. and Decl. of the regions analyzed in this study which correspond to the regions shown in Figure \ref{fig0} and Table \ref{tab2}. 
The reported values are the center of individual apertures placed on those regions.

\begin{deluxetable}{cccc}



\tablecaption{List of R.A. and Decl. of the regions analyzed}

\tablenum{3}

\tablehead{\colhead{Region} & \colhead{ID}\tablenotemark{a} & \colhead{R.A. (deg)} & \colhead{Decl. (deg)} 
} 
\startdata
LyC leaking & & & \\
& 1 & 237.530833 & -78.182511 \\
& 2 & 237.525563 & -78.182751 \\
& 3 & 237.524892 & -78.182810 \\
& 4 & 237.519113 & -78.183177 \\
& 5 & 237.518263 & -78.183252 \\
& 6 & 237.517471 & -78.183350 \\
& 8 & 237.501517 & -78.186255 \\
& 9 & 237.499804 & -78.186744 \\
& 10 & 237.498925 & -78.187074 \\
& 11 & 237.493867 & -78.190767 \\
\hline
Non-leaking & & & \\
& 1 & 237.528421 & -78.182570 \\
& 2 & 237.528054 & -78.182593 \\
& 3 & 237.527658 & -78.182611 \\
& 4 & 237.527267 & -78.182624 \\
& 5 & 237.523850 & -78.182896 \\
& 6 & 237.523546 & -78.182930 \\
& 7 & 237.523188 & -78.182935 \\
& 8 & 237.520983 & -78.183027 \\
& 9 & 237.520621 & -78.183065 \\
& 10 & 237.520146 & -78.183091 \\
& 11 & 237.529838 & -78.182522 \\
& 12 & 237.529383 & -78.182528 \\
& 13 & 237.526692 & -78.182648 \\
& 14 & 237.526317 & -78.182676 \\
& 15 & 237.522508 & -78.182943 \\
& 16 & 237.522054 & -78.182955 \\
& 17 & 237.521646 & -78.182976 \\
& 18 & 237.508696 & -78.184758 \\
& 19 & 237.508238 & -78.184844 \\
& 20 & 237.50750 & -78.184965 \\
& 21 & 237.506454 & -78.185157 \\
& 22 & 237.504383 & -78.185599 \\
& 23 & 237.502675 & -78.186040 \\
& 24 & 237.502267 & -78.186133 \\
\hline
\hline
\enddata

\tablenotetext{a}{ID corresponds to that of Table \ref{tab2}.}


\label{tab3}
\end{deluxetable}

\begin{figure*}
\includegraphics[width=\linewidth]{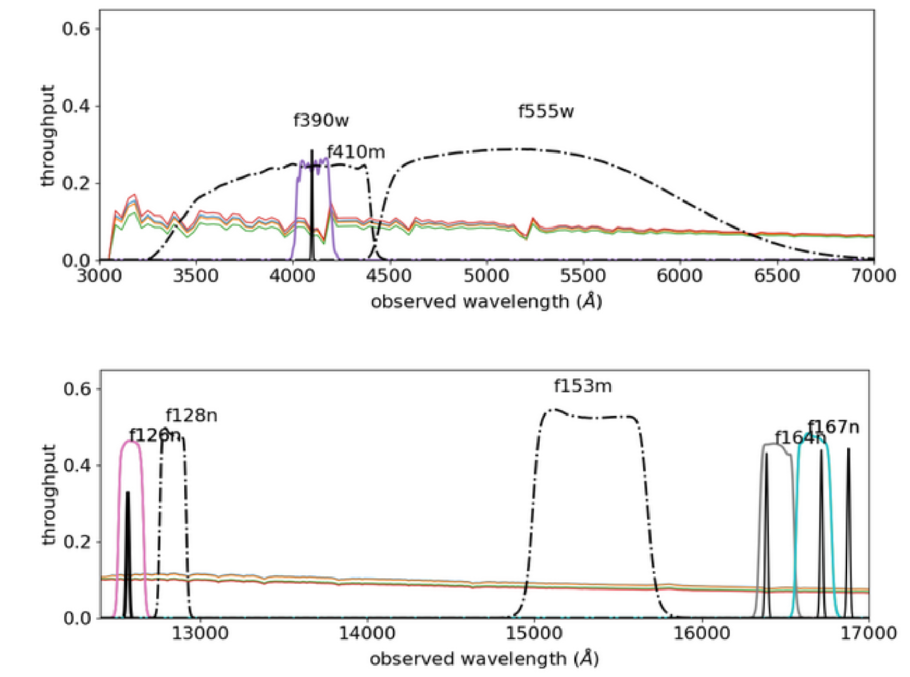}
\caption{The HST filters and the photometric SED fits to subtract off the underlying continuum flux from the narrow-band emission lines (Section \ref{sec:Sec_2_5_emission line measurements}).
On-band filters are shown with solid curves; off-band continuum filters are plotted as dot-dashed curves.
The \textit{y}-axis is the filter throughput. The best-fit spectra used to determine the continuum fits are overplotted as solid lines, with arbitrary scaling (different in each plot, so as not to obscure the filter curves) in $f_{\lambda}$ (units of $\rm{erg \ s^{-1} \ cm^{-2}} \ \angstrom^{\rm -1}$).
Each of the solid lines corresponds to each of the multiple images of the whole galaxy contained in Arcs 1 and 2.
Top panel: the emission line \lya\ is overplotted, with arbitrary flux.
Bottom panel: the following emission lines are overplotted, with arbitrary flux: the \OII\ doublet which falls in the F126N filter; \Hb\ which falls in the F164N filter, \OIII\ $\lambda$4959 which falls in the F1647 filter, and \OIII\ $\lambda$5007 which falls just outside the F167N filter.
\label{fig8}}
\end{figure*}

\end{document}